\documentstyle[preprint,aps]{revtex}
\def\lessim{\mathrel {\vcenter {\baselineskip 0pt \kern 0pt
\hbox{$<$} \kern 0pt \hbox{$\sim$} }}}
\def\gessim{\mathrel {\vcenter {\baselineskip 0pt \kern 0pt
\hbox{$>$} \kern 0pt \hbox{$\sim$} }}}
\def\circ{0}
\def\Pt{$P_T$}
\def\GeVc{GeV$\!/c$}
\def\GeVcc{GeV$\!/c^2$}
\def\Bu{$B^+$}
\def\Bd{$B^\circ$}
\def\Bs{$B_s^\circ$}
\def\mJpsi{J/\psi}
\def\Jpsi{$\mJpsi$}
\def\pbarp{$p\bar{p}$}
\def\Kshort{$K^\circ_s$}
\def\mKshort{K^{\circ}_s}
\def\mKstarplus{K^{\ast}(892)^+}
\def\mKstarzero{K^{\ast}(892)^\circ}
\def\mBR{{\cal B}}
\def\BR{$\mBR$}

\def\invpb{${\rm pb^{-1}}$}

\def\etal{{\it et al.}}
\begin{document}
\title{
\begin{flushright}
FERMILAB-PUB-96/119-E \\
%CDF/PUB/BOTTOM/PUBLIC/3609 \\
%Version 3.0 \\
%
%	0.97  has all prior comments synthesised and was sent to godparents
%	0.971 has Manfred's comments from Mar 4.
%	0.98  has toned down theory discussion and new fb
%	0.981 has comments from GPs, George and Andreas folded in.
%	0.982 has my additional edits...
%	0.99  has George's and my "final" comments
%	1.00  has Manfred's last comments -- this goes out.
%	1.50  has PKS's first pass at integration of comments.
%	1.60  has final corrections before local distribution.
%	1.61  has Ray's comments and some of Manfred's incorporated.
%	1.90  has comments after last godparent committee
%	2.00  has corrected efficiency numbers.
%	2.1   Has some of George's modifications
%	2.2   Has Andreas's corrections also.
%	3.0   Has final changes after review with Manfred and George.
%
\end{flushright}
Ratios of 
bottom meson branching fractions involving ${\bf {\it J}/\psi}$\ mesons
and determination of 
$b$~quark fragmentation fractions} 
\author{
\font\eightit=cmti8
\def\r#1{\ignorespaces $^{#1}$}
\hfilneg
\begin{sloppypar}
\noindent
F.~Abe,\r {14} H.~Akimoto,\r {32}
A.~Akopian,\r {27} M.~G.~Albrow,\r 7 S.~R.~Amendolia,\r {23} 
D.~Amidei,\r {17} J.~Antos,\r {29} C.~Anway-Wiese,\r 4 S.~Aota,\r {32}
G.~Apollinari,\r {27}
% ...
\end{sloppypar}

\vskip .025in
\begin{center}
(CDF Collaboration)
\end{center}

\vskip .025in
\begin{center}
\r 1  {\eightit Argonne National Laboratory, Argonne, Illinois 60439} \\
\r 2  {\eightit Istituto Nazionale di Fisica Nucleare, University of Bologna,
I-40126 Bologna, Italy} \\
\r 3  {\eightit Brandeis University, Waltham, Massachusetts 02254} \\
\r 4  {\eightit University of California at Los Angeles, Los 
Angeles, California  90024} \\  
\r 5  {\eightit University of Chicago, Chicago, Illinois 60637} \\
\r 6  {\eightit Duke University, Durham, North Carolina  27708} \\
\r 7  {\eightit Fermi National Accelerator Laboratory, Batavia, Illinois 
60510} \\
\r 8  {\eightit Laboratori Nazionali di Frascati, Istituto Nazionale di Fisica
               Nucleare, I-00044 Frascati, Italy} \\
\r 9  {\eightit Harvard University, Cambridge, Massachusetts 02138} \\
\r {10} {\eightit Hiroshima University, Higashi-Hiroshima 724, Japan} \\
\r {11} {\eightit University of Illinois, Urbana, Illinois 61801} \\
\r {12} {\eightit Institute of Particle Physics, McGill University, Montreal 
H3A 2T8, and University of Toronto,\\ Toronto M5S 1A7, Canada} \\
\r {13} {\eightit The Johns Hopkins University, Baltimore, Maryland 21218} \\
\r {14} {\eightit National Laboratory for High Energy Physics (KEK), Tsukuba, 
Ibaraki 305, Japan} \\
\r {15} {\eightit Ernest Orland Lawrence Berkeley Laboratory, Berkeley, 
California 94720} \\
\r {16} {\eightit Massachusetts Institute of Technology, Cambridge,
Massachusetts  02139} \\   
\r {17} {\eightit University of Michigan, Ann Arbor, Michigan 48109} \\
\r {18} {\eightit Michigan State University, East Lansing, Michigan  48824} \\
\r {19} {\eightit University of New Mexico, Albuquerque, New Mexico 87131} \\
\r {20} {\eightit Osaka City University, Osaka 588, Japan} \\
\r {21} {\eightit Universita di Padova, Istituto Nazionale di Fisica 
          Nucleare, Sezione di Padova, I-35131 Padova, Italy} \\
\r {22} {\eightit University of Pennsylvania, Philadelphia, 
        Pennsylvania 19104} \\   
\r {23} {\eightit Istituto Nazionale di Fisica Nucleare, University and Scuola
               Normale Superiore of Pisa, I-56100 Pisa, Italy} \\
\r {24} {\eightit University of Pittsburgh, Pittsburgh, Pennsylvania 15260} \\
\r {25} {\eightit Purdue University, West Lafayette, Indiana 47907} \\
\r {26} {\eightit University of Rochester, Rochester, New York 14627} \\
\r {27} {\eightit Rockefeller University, New York, New York 10021} \\
\r {28} {\eightit Rutgers University, Piscataway, New Jersey 08854} \\
\r {29} {\eightit Academia Sinica, Taipei, Taiwan 11529, Republic of China} \\
\r {30} {\eightit Texas A\&M University, College Station, Texas 77843} \\
\r {31} {\eightit Texas Tech University, Lubbock, Texas 79409} \\
\r {32} {\eightit University of Tsukuba, Tsukuba, Ibaraki 305, Japan} \\
\r {33} {\eightit Tufts University, Medford, Massachusetts 02155} \\
\r {34} {\eightit University of Wisconsin, Madison, Wisconsin 53706} \\
\r {35} {\eightit Yale University, New Haven, Connecticut 06511} \\
\end{center}

}
\draft
\address{}
\date{June 3, 1996}
\maketitle
%
%  Now the abstract...
%
\begin{abstract}
We report a measurement of the ratios of the decay rates
of the \Bu, \Bd\ and \Bs\ mesons into exclusive final
states containing a \Jpsi\ meson.
The final states were selected from 19.6~\invpb\ of \pbarp\ collisions
recorded by the Collider Detector at Fermilab.
These data are interpreted to determine the
$b$\ quark fragmentation fractions $f_u$, $f_d$\ and $f_s$.
%We find that $f_u=0.39\pm0.04$, $f_d=0.39\pm0.04$\ and $f_s=0.13\pm0.04$.
We also determine the 
branching fractions for the decay modes
$B^+ \rightarrow J/\psi K^+$,
$B^+ \rightarrow J/\psi K^{\ast}(892)^+$,
$B^\circ \rightarrow J/\psi K^\circ$,
$B^\circ \rightarrow J/\psi K^{\ast}(892)^0$\ and
$B_s^\circ \rightarrow J/\psi\phi(1020)$. 
We discuss the
implications of these measurements to $B$\ meson decay models.
\end{abstract}

% body of paper here
%
\section{Introduction}
The bound states of bottom quarks provide a laboratory in which we
can investigate the behavior of the strong force (Quantum  Chromodynamics or
QCD) and the electroweak interaction\cite{ref: EWK Reference}. 
The lowest-lying bound states are the pseudoscalar mesons (\Bu, \Bd\ and \Bs)
formed by one bottom
anti-quark bound to one of the three lightest quarks ($u$, $d$\ and $s$,
respectively).
The branching fractions (\BR) of these mesons into final
states consisting of only hadrons (the fully hadronic decays) have been studied
theoretically and have been shown to yield insights into the 
interactions that take place between a quark and anti-quark pair at short
distance scales 
\cite{ref: HQET,ref: Deandra et al,ref: Kramer and Palmer I}.

Experimental studies of bottom meson hadronic decays have been
limited by their relatively small branching fractions (typically
$10^{-2}$\ to
$10^{-3}$) and the difficulty of detecting the final states.  
The most
precise branching fraction
measurements have been made at $e^+e^-$\ colliders,
where the \Bu\ and \Bd\ mesons are pair-produced at 
threshold \cite{ref: CLEO results,ref: ARGUS results}. 
Bottom hadrons are produced copiously in high-energy proton-antiproton
collisions \cite{ref: CDF mu Dzero,ref: Dzero b,ref: b Quark sigma}\ and 
so it is possible to measure 
the branching fractions of bottom
mesons into those fully hadronic final states that have distinctive final
state topologies. 
We report a study of
the branching fractions of bottom mesons into final states consisting of a
\Jpsi\ meson and a light quark meson, using the Collider Detector at
Fermilab (CDF).  The data set consists of 19.6 \invpb\ of 1.8 TeV $p\bar{p}$\
collisions produced by the Fermilab Tevatron Collider. 
This work extends an earlier analysis of the same 
dataset \cite{ref: J/psi Kshort BR}\
by incorporating an additional final state, $J/\psi\phi(1020)$,
 estimating the fragmentation fractions
of $B$\ hadrons and providing improved measurements of the branching 
fractions of \Bu\ and \Bd\ mesons.
Throughout this paper, references to a specific decay mode imply the
charge conjugate mode as well.

We have focused our study on the bottom meson decay
modes that yield a \Jpsi\ meson that subsequently
decays to a $\mu^+\mu^-$\ final state. This results in a signature that
we can readily identify using the CDF trigger system and provides the
necessary rejection of other background processes.  We have measured
the observed cross sections times branching fractions for the
channels
\begin{eqnarray}
B^+ &\rightarrow& J/\psi K^+, 
\label{eq: jpsi k} \\
B^\circ &\rightarrow& J/\psi K^\circ, \\
B^+ &\rightarrow& J/\psi K^{\ast}(892)^+, \\
B^\circ &\rightarrow& J/\psi K^{\ast}(892)^0, \\
B_s^\circ &\rightarrow& J/\psi \phi(1020).
\label{eq: jpsi phi}
\end{eqnarray}
The observed cross section for the decay mode 
$ B^+ \rightarrow J/\psi K^+$\ 
can be decomposed into the form
\begin{eqnarray}
\sigma_{\rm obs} = \sigma(p\bar{p}\rightarrow \bar{b})
                    f_u
                    \mBR(B^+ \rightarrow J/\psi K^+)\epsilon^{K^+},
\end{eqnarray}
and similar forms can be written for the other decays.
Here, 
$\sigma(p\bar{p}\rightarrow \bar{b})$\ is the bottom anti-quark production
cross section and $f_u$\ is the probability 
that the fragmentation of 
a $b$\ anti-quark will result in a \Bu\ meson.
In a similar way, we define $f_d$\ and $f_s$\ to be the probabilities
of a $b$\ anti-quark to hadronize and form a \Bd\ and \Bs\ meson,
respectively.  We will refer to these probabilities as fragmentation
fractions and explicitly include in this fraction
contributions from decays of heavier
$B$\ hadrons into final states containing a \Bu, \Bd\ or \Bs\ meson.
The expression $\mBR(B^+ \rightarrow J/\psi K^+)$\ represents the
branching fraction for this decay mode of the \Bu\ meson and 
$\epsilon^{K^+}$\ is the efficiency of detecting the 
$J/\psi K^+$\ final state. 

The fragmentation fractions into the different $B$\ meson states are
not well known.
If one assumes that these fractions are independent of the flavor and 
energy of the
quark initiating the hadronization process,  
then measurements of strange meson production in 
light quark fragmentation provide the most accurate 
estimates of $f_s$\ \cite{ref: K production}.
Only one measurement 
of $f_s$\ in the bottom meson system exists \cite{ref: Delphi fs}\
and it has large uncertainties.
No measurements have been made of $f_u$\ or $f_d$\ in the bottom meson
system, despite the importance of these in many $B$\ hadron 
branching fraction, lifetime and mixing studies
\cite{ref: PDG}.

The bottom quark production cross section
$\sigma(p\bar{p}\rightarrow \bar{b})$\ is not precisely known; the best
measurements to date have uncertainties of order $\pm20$\%\
\cite{ref: b Quark sigma}.  
The efficiency $\epsilon^{K^+}$\
depends on an understanding of the
$b$~quark production properties and detector acceptance.
We therefore present our measurements in the
form of ratios of branching fractions in order to avoid introducing
additional uncertainties
due to the $b$\ quark production cross section and 
the final state detection efficiency.  We then compare these ratios
directly with phenomenological predictions of the relative $B$\ meson
branching fractions.
We also use our data to estimate the fragmentation fractions
$f_u$, $f_d$\ and $f_s$.

We have organized this report as follows.  In Section II, we describe the
data selection and procedures that we followed to reconstruct the 
five decay modes.  We present in Section III a study of the relative
sizes of resonant and non-resonant $K\pi$\ and $KK$\ contributions to
the $B$~meson final states.  We describe the procedure used to determine
the acceptance and efficiency corrections for each decay mode in Section
IV.  In Section V, we present the results of this study and conclude
in Section VI.

\section{Data Collection and Selection}
\subsection{The CDF Detector}
CDF is a multi-purpose detector designed to study
high-energy \pbarp\ collisions produced by the Fermilab Tevatron
Collider. 
It surrounds the interaction point
with three charged particle tracking detectors immersed in a 1.4~T
solenoidal magnetic field.
The tracking system is contained within a hermetic calorimeter system that
measures the energy flow of charged and neutral particles.  Charged particle
detectors outside the calorimeter are used to identify muon candidates.
The detector has a coordinate system with the $z$\ axis along the
proton beam direction.  The polar angle $\theta$\ is defined relative to 
the $z$\ axis, $r$\ is the radius from this axis
and $\phi$\ is the azimuthal angle. 
Pseudorapidity is defined as $\eta\equiv-\ln\tan(\theta/2)$.

The innermost tracking device is a silicon microstrip detector (SVX)
located in the region between 3.0 and 7.9 cm in radius from the beam axis.
This is followed by a set of time projection chambers (VTX) that measure
charged particle trajectories out to a radius of 22 cm.  An 84-layer drift
chamber (CTC) measures the particle trajectories in the region between 30
and 130 cm in radius from the beam.  This tracking system has high
efficiency for detecting charged particles with momentum transverse to the beam
$P_T>0.35$\ \GeVc\ and $|\eta|\lessim1.1$, and
the CTC and SVX together measure charged particle transverse
 momenta with a precision of 
$\sigma_{P_T} \sim [(0.0066 {P_T})^2 + (0.0009 P_T^2)^2]^{1/2}$\ 
(with \Pt\ in units of \GeVc).  

The muon detection  system consists of
4 layers of planar drift chambers separated from the
interaction point by $\sim5$\ interaction lengths of material.  
Additional 4 layers of chambers are located outside the magnet return
yoke (corresponding to 4 interaction lengths of material) 
in the central pseudorapidity region $|\eta|<0.7$\ to reduce
the probability of misidentifying penetrating hadrons as muon candidates.
An additional set of chambers is located in the pseudorapidity interval
$0.7<|\eta|<1.0$\ to extend the acceptance of the muon system.
The muon system is
capable of detecting muons with $P_T \gessim 1.4$\ \GeVc\ in a
pseudorapidity interval $|\eta| < 1.0$.
These and other elements of the CDF detector are described in more detail
elsewhere \cite{ref: CDF Detector}.

\subsection{The $J/\psi$\ Selection}
We selected the \Jpsi\ final state using a three-level trigger system that
identified collisions with two muon candidates.
The first level trigger required that there be two track candidates observed
in the muon system.  The level one trigger track efficiency 
rises from $\sim40$\% at 
$P_T=1.5$~\GeVc\ to $\sim93$\%\ for muons with $P_T>3$~\GeVc.
The second level trigger requires the detection of a charged track in the
CTC using the Central Fast Track processor (CFT), which performs a partial
reconstruction of all charged tracks above a transverse 
momentum of $\sim2.5$~\GeVc.
The CTC track is required to match within $15^\circ$\ in $\phi$\ of the
muon candidate.
The CFT efficiency rises from 40\%\ at a muon
$P_T=2.6$~\GeVc\ to $\sim94$\%\ for $P_T>3.1$~\GeVc.  
The third level trigger requires that two reconstructed CTC
tracks match with two tracks in the muon chambers and that the dimuon invariant
mass  be between 2.8 and 3.4~\GeVcc.  
The efficiency of the level three
trigger requirement is $(97\pm2)$\%\ for \Jpsi\ candidates. 
There are $2.06\times10^5$\ dimuon candidate events that passed the level three 
trigger requirements.

\begin{figure}
\vspace*{5in}
\vskip 3in
\hbox{
\includegraphics{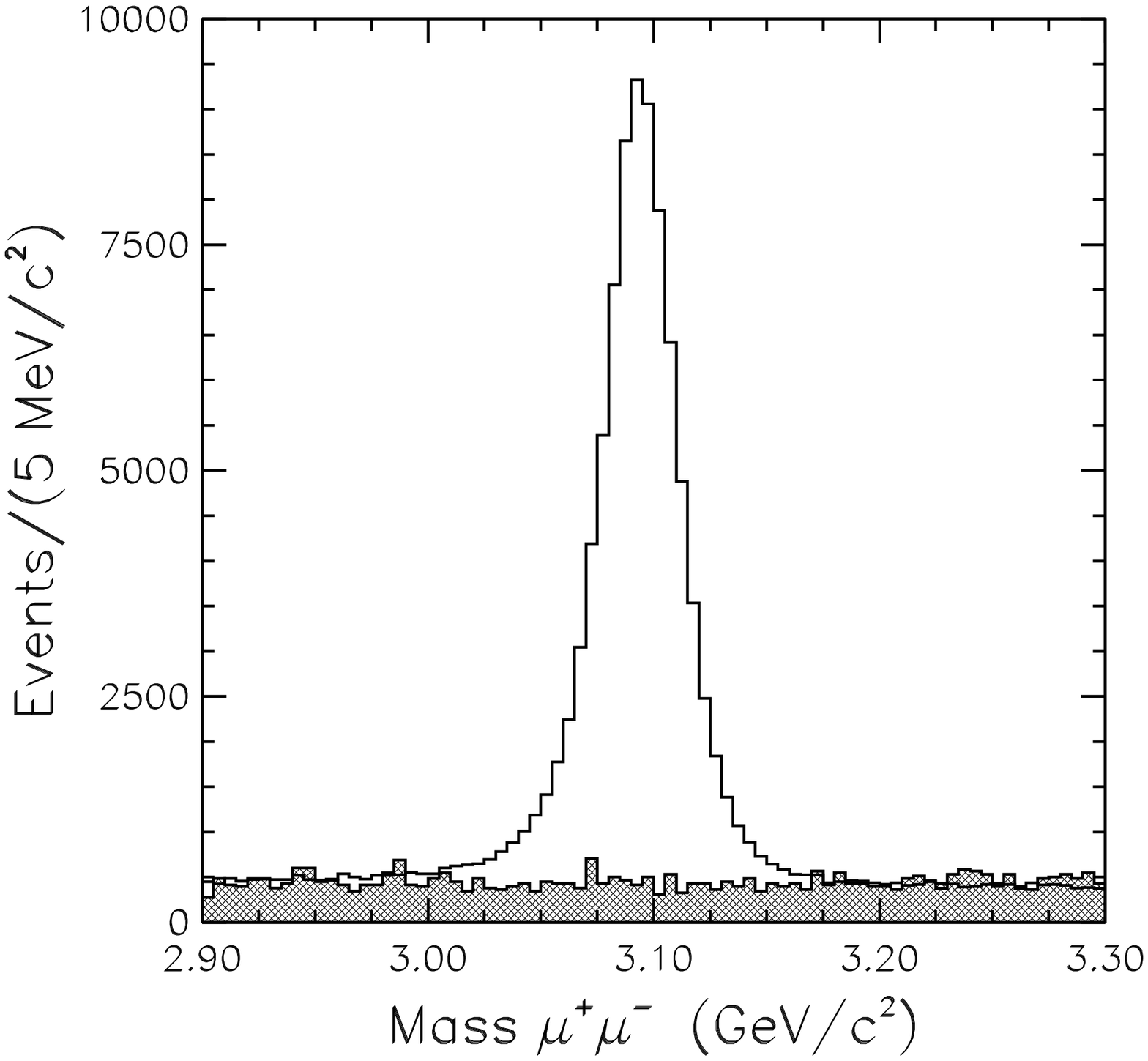}
}
\vskip -1in
\caption{The dimuon invariant mass distribution for the inclusive \Jpsi\
event sample.  The shaded distribution is for same-sign
dimuon candidates. 
}
\label{fig:  invariant masses}
\end{figure}

These events were further selected to identify a clean sample of \Jpsi\ 
candidates.
We required that each muon candidate have a CTC
track candidate with $P_T>1.4$\ \GeVc. This track, when extrapolated to the muon
chambers, was required to match within 3 standard deviations of the 
extrapolation and measurement uncertainties 
with a muon track in the transverse plane ($r$-$\phi$)
and  along the beam axis direction.  
The two muon candidates were required to have opposite charges.
We performed a least-squares fit of the two muon
candidate tracks under the constraint that the two tracks come from 
a common point (a vertex constraint).
We required the probability of this fit to be greater than 0.01. 
These requirements resulted in a signal of $(7.89\pm0.08)\times10^4$\
\Jpsi\ decays on a background of non-resonant dimuon
candidate events. The dimuon invariant mass distribution for this sample
is shown in  Fig.~\ref{fig:  invariant masses}, along with an estimate
of the background determined using same-charge muon candidate pairs. We
performed an additional fit to the dimuon system, applying a vertex
constraint and requiring that the dimuon invariant mass equal the world
average \Jpsi\ mass of 3.09688~\GeVcc\ \cite{ref: PDG}. The confidence level
of this vertex-plus-mass constrained fit was required to be greater than 0.01.  

\subsection{Reconstruction of Exclusive Decays}
%
%  First J/psi K+
%
\subsubsection{The $B^+ \rightarrow J/\psi K^+$\ Channel}
We reconstructed the exclusive decay modes listed 
in Eqs.~\ref{eq: jpsi k}-\ref{eq: jpsi phi}\  by
forming charged particle combinations with the \Jpsi\ candidate.  For the
decay channel $B^+ \rightarrow J/\psi K^+$, we 
considered every charged particle with $P_T>1.5$~\GeVc\ as a $K^+$\ 
candidate and required the resulting \Bu\ candidate to have $P_T >
8.0$\ \GeVc. 
A least-squares fit was performed on the three charged
tracks forming the $J/\psi K^+$\ candidate by constraining the three tracks to
come from a common vertex, the invariant mass of the dimuon system to
equal the world average \Jpsi\ mass, and the
flight path of the $B^+$\ candidate to be parallel to its momentum
vector in the transverse plane (a 2-dimensional pointing constraint).  
The confidence level of this least-squares fit had to exceed 0.01.
We required the fitted transverse momenta of the muon candidate with
lowest and highest \Pt\ to be greater than 1.8 and 2.5~\GeVc, respectively.
This ensured that the muon candidates were likely to pass the dimuon
trigger requirements.
In order to reduce the backgrounds from prompt \Jpsi\ 
production, we required the \Bu\ meson candidate flight path to
be pointing in the same hemisphere as its momentum vector
(in effect requiring the \Bu\ candidate's observed proper decay 
length, $c\tau$, to be positive).
The interaction vertex position was determined by averaging the measured
beam position over a large number of collisions recorded under identical
Tevatron Collider operating conditions.
The $J/\psi K^+$\ invariant mass distribution is shown
in Fig.~\ref{fig: Jpsi K Mass}(a).
We have performed a binned maximum likelihood fit
of this data to a Gaussian lineshape 
and a linear background term and
estimate a \Bu\ signal of $154\pm19$\ events.  The width of the signal
was not constrained in the fit and resulted in a fitted 
mass resolution of $0.015\pm0.002$ \GeVcc,
consistent with our expected detector resolution.

The reconstruction of the other four decay modes was performed with 
similar criteria in order to reduce the systematic uncertainties
resulting from the kinematics of the produced $B$\ mesons and 
selection biases.
Identical requirements were made on the quality and transverse momenta
of the muon candidates, constraints on the fits to the $B$\ decay topologies,
$B$\ meson lifetimes and \Kshort\ lifetimes.
We allowed for small variations in the $B$\ and light quark meson
transverse momentum requirements to optimize the
expected significance for each channel.  
The significance is defined as $N_s/\sqrt{N_s + N_b}$, where $N_s$\ is the
expected number of events determined using a Monte Carlo calculation for
a given integrated luminosity, and $N_b$\ is the extrapolated background rate
under the signal region using the observed $B$\ meson sideband background
levels.
This resulted in only modest
differences in the $P_T$\ requirements from channel to channel, and
did not introduce significant 
systematic uncertainties in our estimation of the $B$~meson detection
efficiency.

%
%  Now J/psi K0
%
\subsubsection{The $B^\circ \rightarrow J/\psi K^\circ$\ Channel}
\begin{figure}
\vspace*{0.2in}
\begin{picture}(432,12)
%\put( 50,0){\bf (a)}
%\put(266,0){\bf (b)}
\end{picture}
\vspace*{1.7in}
\vskip 1in
\hbox{
\includegraphics{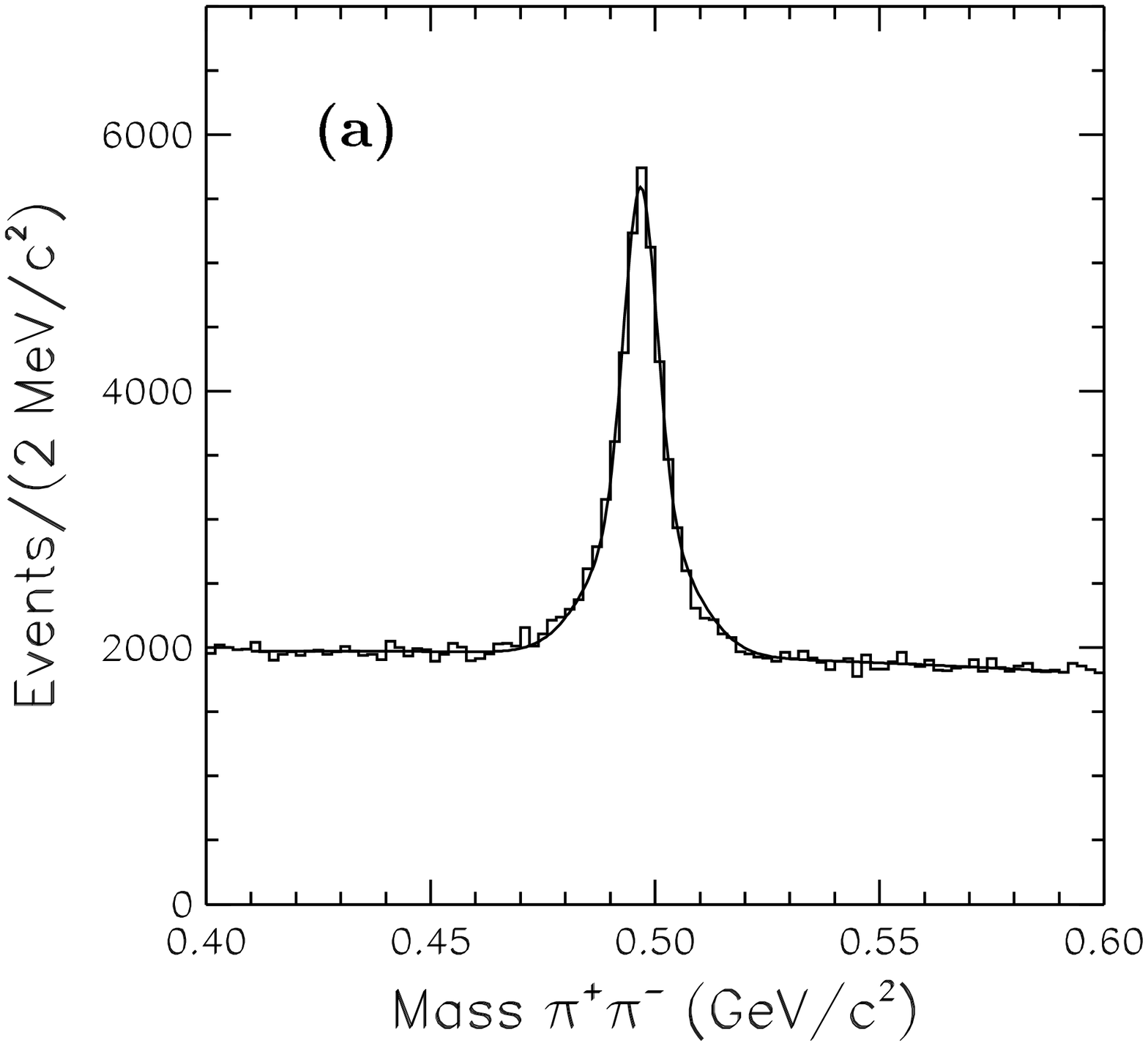}
\hskip 3.0in 
\includegraphics{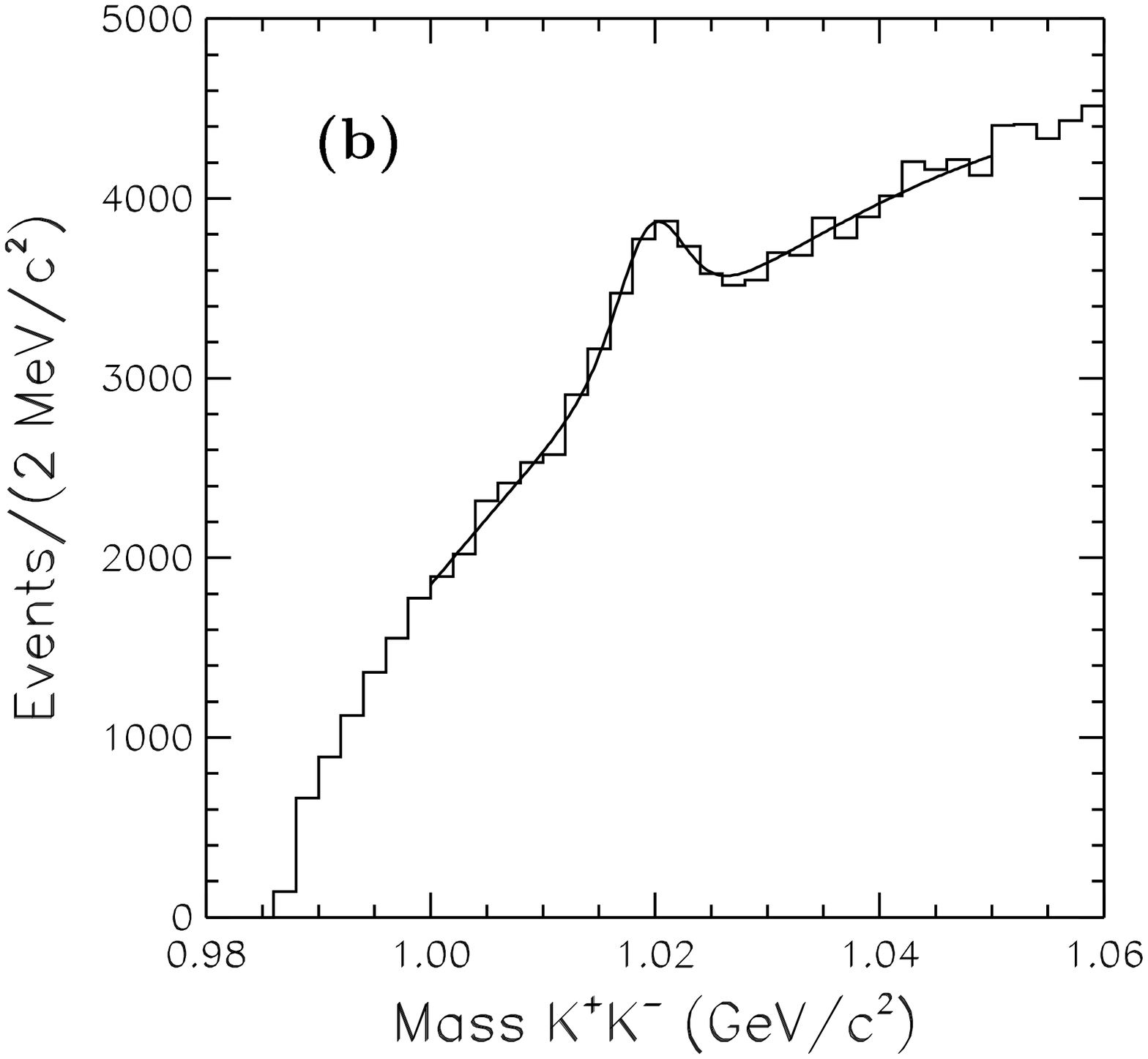}
}
\vspace*{0.0in}
\begin{picture}(432,12)
%\put( 50,15){\bf (c)}
%\put(266,15){\bf (d)}
\end{picture}
\vspace*{2.5in}
\hbox{
\includegraphics{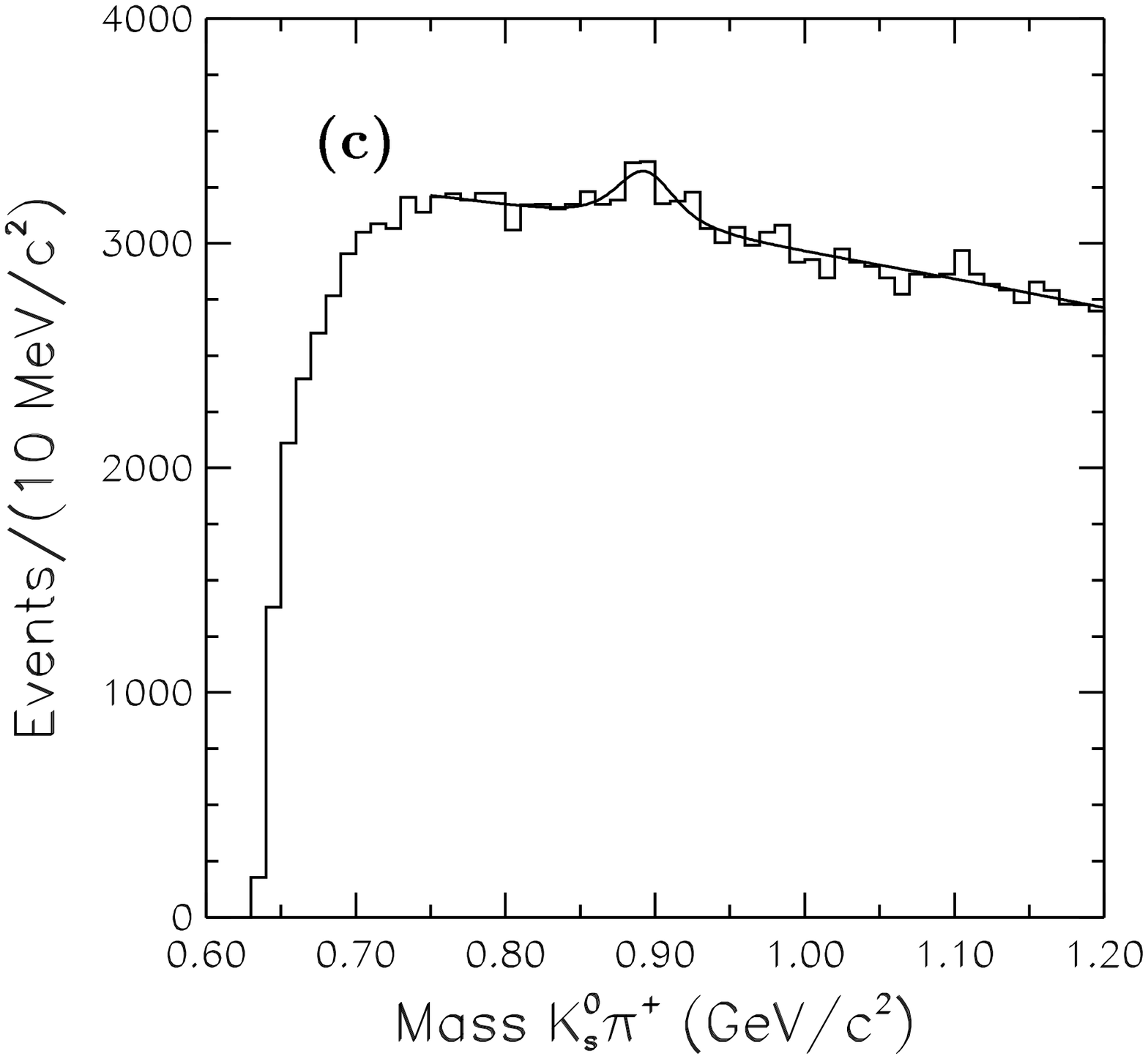}
\hskip 3.0in 
\includegraphics{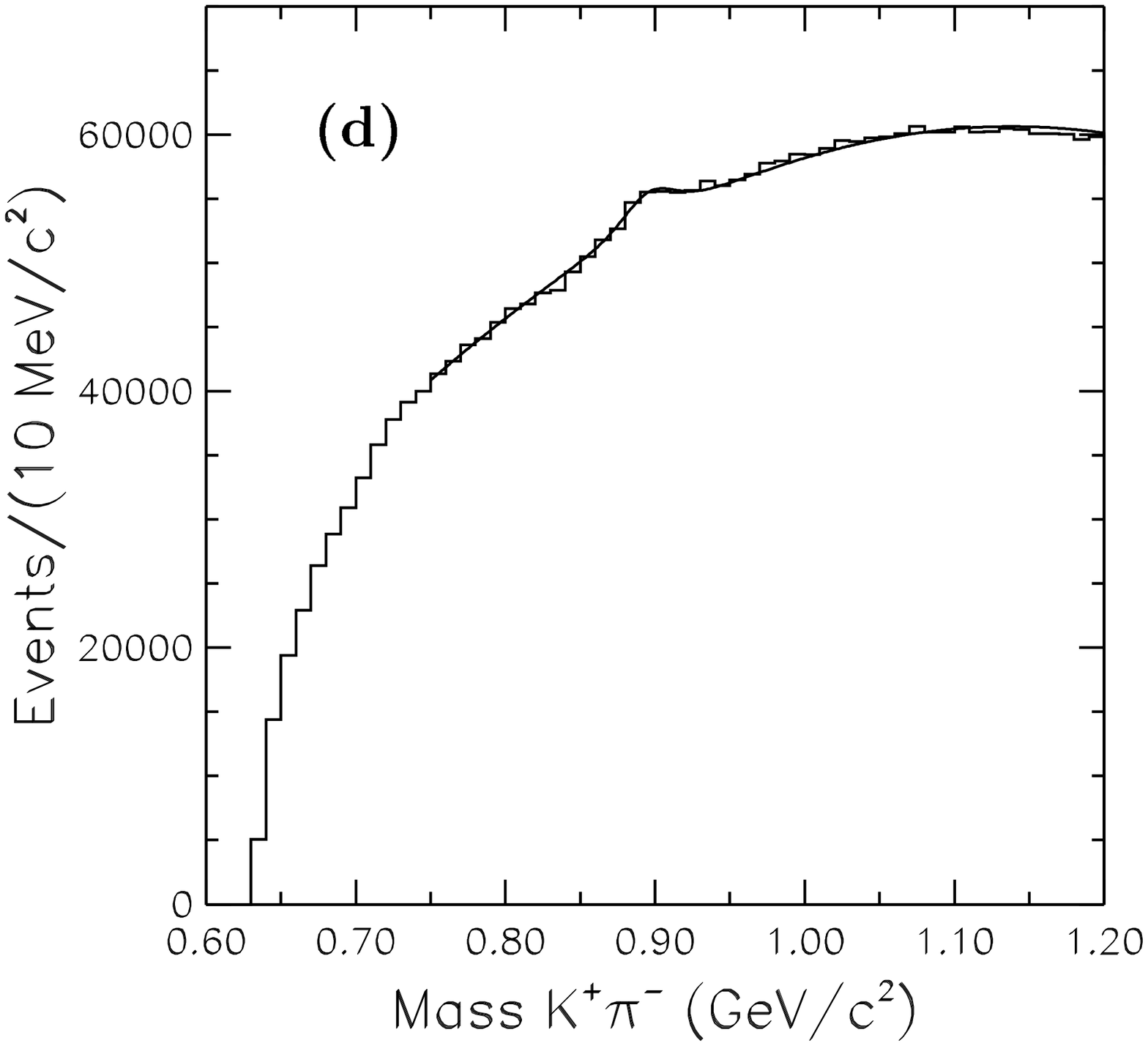}
}
%\hskip 1.5in \special{psfile=prd_plots/prd3.ps hscale=40 vscale=40}
%
\caption{
The $\pi^+\pi^-$\ invariant mass distribution is shown in (a) for
\Kshort\ candidates. 
The $K^+K^-$\ invariant mass distribution 
is shown in (b).
The $\mKshort\pi^+$\ invariant mass distribution is shown in (c).
The $K^+\pi^-$\ invariant mass distribution is shown in (d).
}
\label{fig: inclusive masses}
\end{figure}

The decay mode $B^\circ\rightarrow J/\psi K^\circ$\ was reconstructed
by searching for $K^\circ \rightarrow \mKshort\rightarrow \pi^+\pi^-$\ 
candidates using all pairs of 
oppositely charged particles. 
The daughter pions were required to have $P_T>0.4$~\GeVc.
To fully reconstruct the $B^\circ\rightarrow J/\psi\mKshort$\ decay,
we performed a least-squares fit to the two
pion candidate tracks and two muon candidates, constraining 
each track pair to
come from common points, requiring the momentum vector of the
$\mKshort$\ candidate to point along its flight path (a vertex and pointing
constraint) and placing a \Jpsi\ mass constraint on the 
dimuon system.
We also imposed a mass constraint on the dipion system, constraining the
invariant $\pi^+\pi^-$\ mass to the world average \Kshort\ mass of 
0.4977~\GeVcc.
The confidence level of the
fit had to exceed 0.01.  
To improve the signal-to-noise ratio, we
required the proper decay length of the $K^\circ_s$\
candidate to be larger than 0.1~cm and its transverse momentum to be
greater than 1.5~\GeVc.  
The $\pi^+\pi^-$\ invariant mass distribution for the 
\Kshort\ candidates that satisfy these requirements is illustrated in 
Fig.~\ref{fig: inclusive masses}(a), and shows a \Kshort\ signal
of $(2.56\pm0.05)\times10^4$\ decays
above a large combinatorial background (for illustration,
no mass constraints were imposed in the least-squares fit to the
charged tracks in this figure).  To identify a clean \Bd\ candidate sample,
we required the $J/\psi \mKshort$\ candidates 
to have $P_T>6.0$~\GeVc\
and the candidates to have a $c\tau$\ greater
than zero to reduce the combinatorial backgrounds from prompt $J/\psi$\
production.
The invariant
$J/\psi K^\circ_s$\ mass distribution for these candidates is shown in
Fig.~\ref{fig: Jpsi K Mass}(b).
A fit of this distribution to a Gaussian lineshape and linear background
results in a total signal of $36.9\pm7.3$\
$B^\circ$\ decays.

%
%  Next is J/psi K*+/-
%
\subsubsection{The $B^+ \rightarrow J/\psi \mKstarplus$\ Channel}
We searched for the decay mode 
$B^+ \rightarrow \mJpsi \mKstarplus \rightarrow
J/\psi K^\circ_s \pi^+$\ by  selecting a sample of \Jpsi\ candidate events
containing a
$K^\circ_s$\ candidate. 
The criteria used to identify \Kshort\ candidates for the $J/\psi\mKshort$\
final state were also used for this decay mode.
We required the \Kshort\ candidates to have $P_T>2.0$~\GeVc.
We considered 
all other charged tracks with $P_T>0.4$\ \GeVc\ as $\pi^+$\
candidates and we combined these with the \Kshort\ candidates to form all
possible $\mKstarplus\rightarrow K^\circ_s \pi^+$\ candidates.
The combinatorial backgrounds to the $\mKstarplus$\ decay are 
large, as illustrated in the $\mKshort\pi^+$\ invariant mass distribution 
presented in Fig.~\ref{fig: inclusive masses}(c).
In order to identify a \Bu\ candidate sample,
a least-squares fit similar to that imposed on the
$J/\psi K^\circ_s$\ candidates was performed.
We required that the
confidence level of this fit be greater than 0.01, and that the $c\tau$\ of
the $K^\circ_s$\ candidate be greater than 0.1~cm. 
The transverse
momentum of  the $B^+$\ candidate had to exceed 6.0~\GeVc\ and its
$c\tau$\ had to be positive.
In order to isolate a $K^{\ast}(892)^+$\ resonance, we required 
that the invariant $\mKshort\pi^+$\ mass be within 0.08~\GeVcc\ of the
world average  $K^{\ast}(892)^+$\ mass (0.8916\ \GeVcc) \cite{ref: PDG}.
This results in the $J/\psi \mKshort \pi^+$\ invariant mass distribution
shown in Fig.~\ref{fig: Jpsi K Mass}(c).  A fit of this
distribution to a Gaussian lineshape and linear background
results in an estimated signal of $12.9\pm4.3$\ decays.

%
%  Not the B^\circ -> J/psi K*0
%
\subsubsection{The $B^\circ \rightarrow J/\psi \mKstarzero$\ Channel}
Our data selection to reconstruct the decay 
$B^\circ \rightarrow J/\psi K^{\ast}(892)^0 \rightarrow J/\psi K^+ \pi^-$\
proceeded in a similar manner.  
We formed combinations of all oppositely-charged
track pairs, and fit the four charged tracks
requiring that they come from a common decay point, constraining the
invariant dimuon mass to the world average
\Jpsi\ mass, and requiring that the flight path of the \Bd\ candidate
be parallel to its momentum vector in the transverse plane.  The confidence
level of this fit had to be greater than 0.01 and the \Bd\ candidate
$c\tau$\ had to be positive.   
The combinatorial backgrounds to the $\mKstarzero$\ decay are also
large.  This is illustrated in Fig.~\ref{fig: inclusive masses}(d),
where we show the
$K^+\pi^-$\ invariant mass distribution for events that have 
transverse momentum of the $K^+\pi^-$\ system greater than 2.0~\GeVc.
We defined the \Bd\ candidate sample by requiring
the \Pt\ of the $K^+\pi^-$\ 
system to be greater than 2.0~\GeVc\ and the resulting $J/\psi
K^+\pi^-$\ system to have
$P_T > 8.0$~\GeVc. 
We required the $K^+\pi^-$\ invariant mass to be within 0.08~\GeVcc\ of
the world average $K^{\ast}(892)^0$\ mass (0.8961\ \GeVcc). 
The resulting
$J/\psi K^+\pi^-$\ invariant mass distribution is shown in 
Fig.~\ref{fig: Jpsi K Mass}(d).  

This peak also has contributions from $K^{\ast}(892)^0$\ decays 
where the incorrect
kaon and pion mass assignments yield an invariant $K^+\pi^-$\
mass within the $K^\ast(892)^\circ$\ mass window of $\pm0.08$~\GeVcc.
We used a Monte Carlo calculation, described in 
Section IV, to determine the relative fraction of
such combinations and the shape of the resulting $J/\psi K^+\pi^-$\
invariant mass distribution.  
The signal shape was parametrized by two Gaussian distributions
with the relative width, normalization and position of the second distribution
determined
by a fit to the $J/\psi K^+\pi^-$\ invariant mass distribution predicted
by the Monte Carlo calculation.  
The width of the second Gaussian was fixed to 3.3 times the width of
the first, the normalization of the second was fixed to 0.08 times
that of the first and the mean
of the second Gaussian distribution was offset lower in mass by
0.0023~\GeVcc\ relative to the mean of the first.
This shape was then used in a fit to the observed $J/\psi \mKstarzero$\ 
invariant mass distribution to determine the
number of $B^\circ$\ decays in our data.
This procedure yields a signal of
$95.5\pm14.3$\ 
$B^\circ\rightarrow J/\psi \mKstarzero$\ decays. 

%
%  Finally J/psi phi
%
\subsubsection{The $B_s^\circ \rightarrow J/\psi \phi(1020)$\ Channel}
The search for the decay mode 
$B_s^\circ\rightarrow J/\psi \phi(1020) \rightarrow
J/\psi K^+ K^-$\ was performed by considering as $\phi(1020)\rightarrow
K^+K^-$\ candidates all oppositely-charged track pairs. 
A least-squares fit of the $\mu^+\mu^-K^+K^-$\
candidate system was performed, constraining all four tracks to come from
the same vertex, constraining the dimuon invariant mass to the world average
\Jpsi\ mass, and imposing a 2-dimensional pointing constraint on the
\Bs\ decay.
The confidence level of this fit had to exceed 0.01.
A $\phi(1020)$\ signal of $(4.1\pm0.4)\times10^3$\ events is evident in
this sample, as illustrated in Fig.~\ref{fig: inclusive masses}(b).
The combinatorial background was reduced by requiring the $K^+K^-$\
system to have
$P_T>2.0$~\GeVc, the  $J/\psi K^+ K^-$\ system to have $P_T>6.0$~\GeVc\
and the $c\tau$\ of the \Bs\ candidate system
to be positive.
We defined our $\phi(1020)\rightarrow K^+K^-$\ candidate sample by requiring the
 $K^+K^-$\ invariant mass to be within 0.0100~\GeVcc\ of the world
average
$\phi$\ mass (1.0194~\GeVcc) \cite{ref: PDG}.
This resulted in a sample with the 
$J/\psi K^+K^-$\ invariant mass distribution shown in 
Fig.~\ref{fig: Jpsi K Mass}(e).  A \Bs\ signal is evident on a relatively
small background.  A fit of this distribution to a Gaussian lineshape  
and linear background results in a total signal of
$29.4\pm6.2$\ events.

\section{Resonant and Non-Resonant Decays}
Clear signals for $B$\ meson production and decay are observed in all
five channels.  In the case of the three channels involving a 
$K^\ast(892)$\ or $\phi(1020)$\ resonance in the final state, the
estimated number of $B$\ candidate events includes 
resonant and non-resonant contributions in the
final state.  We searched for evidence of a non-resonant $K\pi$\ or $KK$\
contribution to the $B$~meson signals
by placing invariant mass cuts on the $B$~candidate, removing
the invariant mass cuts on the two-meson systems
and examining the $\mKshort\pi^-$, $K^+\pi^-$\ and $K^+K^-$\ invariant 
mass distributions. 
In order to account for non-$B$\ background in the
two-body mass distributions, we  defined $B$\ mass sideband regions 
for the three samples, normalized to the estimated number of non-$B$\
events as determined from the $B$\ invariant mass distributions.
The signal and sideband regions are described in
Table~\ref{tab: sideband regions}.
By allowing for a non-resonant contribution to the $B$\ decay rate,
we will directly estimate the rate of resonant decays without having to
assume that the rate of non-resonant decays is negligible.

\begin{figure}
\begin{picture}(432,12)
%\put( 45, -10){\bf (a)}
%\put(266, -10){\bf (b)}
\end{picture}
\vspace*{2.5in}
\vskip 0.7in
\hbox{
\includegraphics{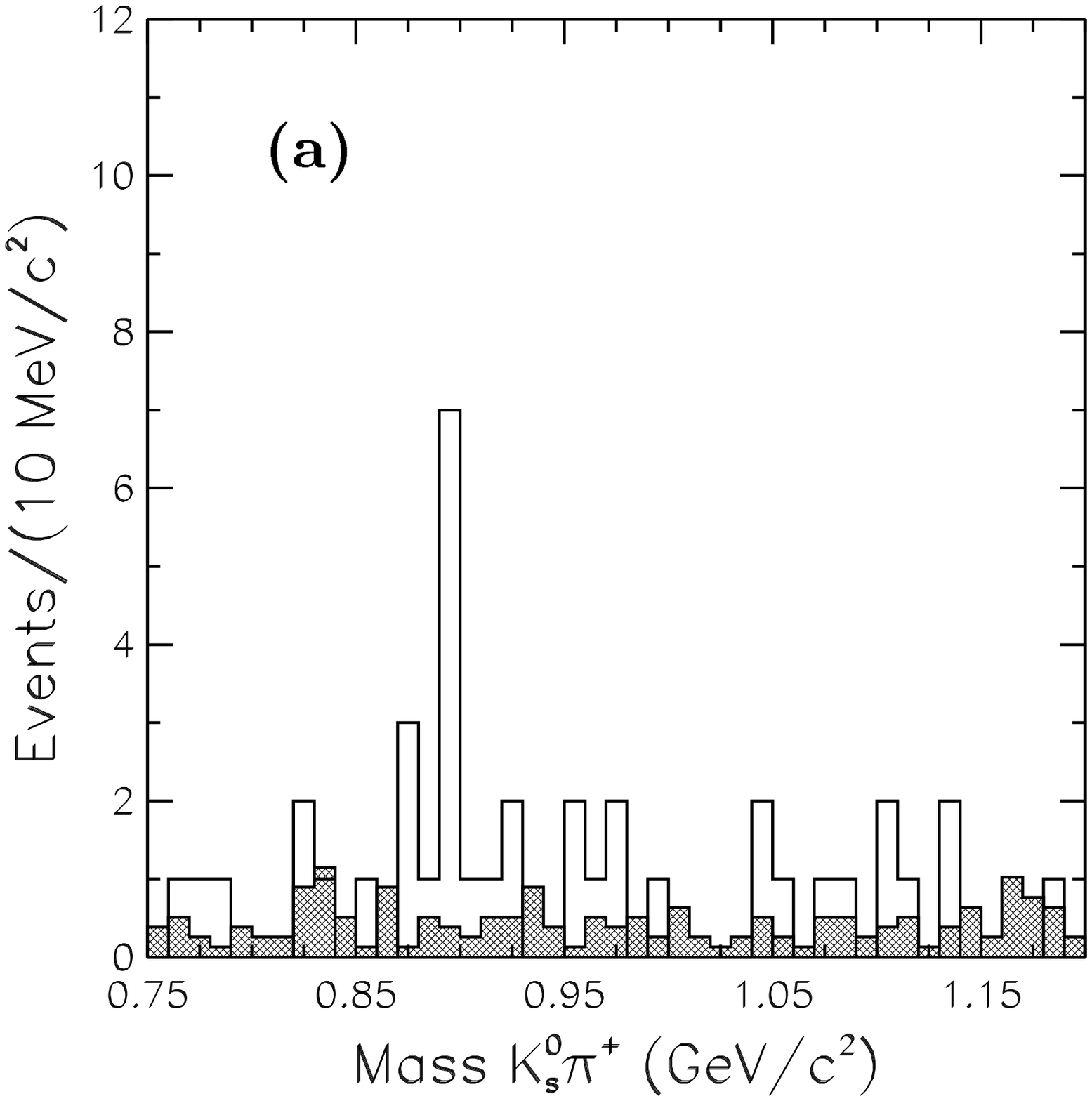}
\hskip 3.0in \includegraphics{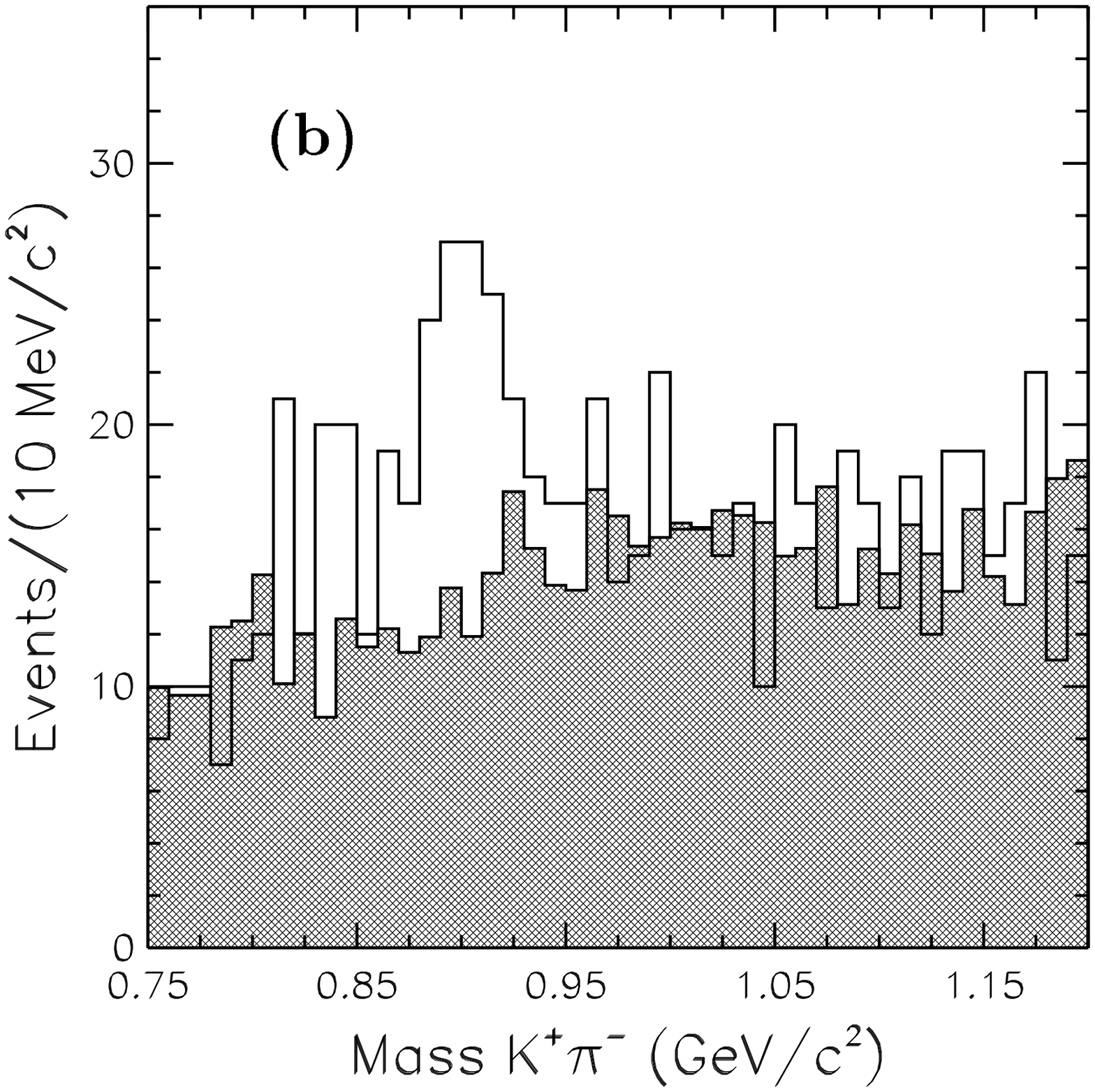}
}
\begin{picture}(432,12)
%\put(150,25){\bf (c)}
\end{picture}
\vspace*{2.7in}
\hbox{
\hskip 1.5in \includegraphics{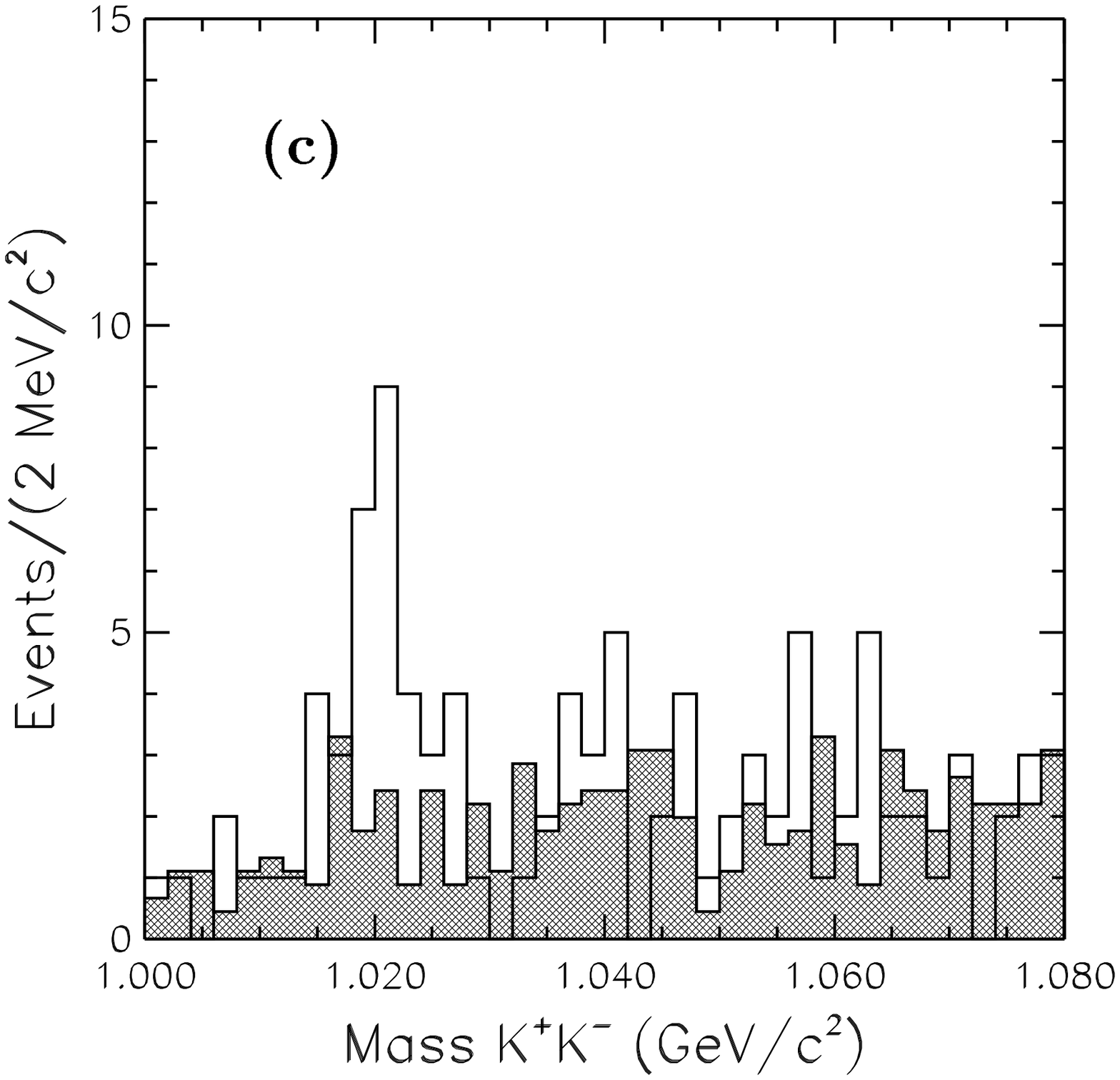}
}
\caption{The $\mKshort\pi^+$\ invariant mass distributions for the 
$B^+$\ signal and sideband regions are shown in (a).  The $B^+$\ signal
region is represented by the unshaded histogram.  The $B^+$\ sideband
region, normalized to the non-$B^+$\ background in the $B^+$\ signal
region, is the shaded distribution.
The corresponding distributions for the $K^+\pi^-$\ and $K^+K^-$\
invariant mass are shown in (b) and (c), respectively.
}
\label{fig: resonances}
\end{figure}

The $\mKshort\pi^+$, $K^+\pi^-$\ and $K^+K^-$\ invariant mass distributions
are illustrated in Fig.~\ref{fig: resonances}\ for the $B$\ signal
and sideband regions (the shaded distributions are from candidates in the
$B$\ sideband regions normalized to the background under the $B$\ meson 
peak in the signal region). One
sees from these distributions resonant signals for the
$K^\ast(892)$\ and $\phi(1020)$\ with no significant non-resonant contributions
above the non-$B$\ backgrounds.  We quantified the amount of resonance
production associated with the $B$\ signals by performing binned maximum
likelihood fits of the two-body invariant mass distributions to
Breit-Wigner lineshapes convoluted with detector
resolution, using the normalized $B$\ sideband distributions to
model the shape and size of the background under the two-body resonance 
signals.
The resulting numbers of observed events, $N_{sb}$, are 
listed in Table~\ref{tab: sideband regions}.
As a cross-check, we also estimated the number of signal events 
by fitting the resonance signals to 
Breit-Wigner lineshapes convoluted with detector
resolution and background shapes described by
second-order polynomial functions.  The resulting event rates, $N_{fit}$,
are listed in Table~\ref{tab: sideband regions}\ and are 
consistent with $N_{sb}$.

Under the assumption that there are no non-resonant 
decays, we can also estimate the strength of the two-body resonant decay
by correcting the observed $B$\ rates determined from the fits to
the $J/\psi K\pi$\ and $J/\psi KK$\ invariant mass distributions 
in Fig.~\ref{fig: Jpsi K Mass}(c)-(e) for
the loss in efficiency due to the $K\pi$\ and $KK$\ mass cuts. 
The presence of non-resonant $K\pi$\ or $KK$\ decays would result in 
corrected $B$\ decay rates systematically larger than those determined by
$N_{sb}$\ or $N_{fit}$.   
The mass cut efficiencies have been estimated using
a Monte Carlo calculation to be 0.80 and 0.86 for the $K\pi$\ and $KK$\ 
mass window cuts, respectively.
The resulting $B$\ meson
rates, $N_{win}$, are 
listed in Table~\ref{tab: sideband regions}.  We see no 
significant difference in the rates estimated by these 
three methods.  We therefore
conclude that we do not observe a significant non-resonant
$B \rightarrow J/\psi K\pi$\  or $B_s^\circ\rightarrow J/\psi KK$\ decay
mode.

For the subsequent analysis, 
we choose $N_{sb}$\ as the best estimate of the rate of resonant 
production as it is least biased by potential contributions from 
non-resonant production.
In addition, we have investigated the possibility that kinematic reflections
of other $B$~hadron decay modes could enhance our observed event yields, and
have excluded such contributions.

\section{Efficiency Corrections}
We estimate the relative reconstruction efficiency for each $B$\
meson decay mode to convert the observed number of $B$\ events into ratios
of branching fractions.  
We write the efficiencies for reconstructing $B$\ mesons as
\begin{eqnarray}
\epsilon^{K^+} &=&
\epsilon_{J/\psi}
\times \epsilon^{K^+}_{geom}
\times \epsilon^{K^+}_{c\tau}
\times \epsilon_{K} \\
\epsilon^{K^\circ} &=&
\epsilon_{J/\psi}
\times \epsilon^{K^\circ}_{geom}
\times \epsilon^{K^\circ}_{c\tau}
\times \epsilon_{K_s} \\
\epsilon^{K^{\ast+}} &=&
\epsilon_{J/\psi}
\times \epsilon^{K^{\ast+}}_{geom}
\times \epsilon^{K^{\ast+}}_{c\tau}
\times \epsilon_{K_s} 
\times \epsilon_{\pi}\\
\epsilon^{K^{\ast\circ}} &=&
\epsilon_{J/\psi}
\times \epsilon^{K^{\ast\circ}}_{geom}
\times \epsilon^{K^{\ast\circ}}_{c\tau}
\times \epsilon_{K}
\times \epsilon_{\pi} \\
\epsilon^{\phi} &=&
\epsilon_{J/\psi}
\times \epsilon^{\phi}_{geom}
\times \epsilon^{\phi}_{c\tau}
\times \epsilon_{\phi} ,
\end{eqnarray}
where we show the common contributions.
The quantity
$\epsilon_{J/\psi}$\ is the efficiency for triggering and
reconstructing the $J/\psi\rightarrow \mu^+\mu^-$\
decay.  It is common to all decay modes.
This also includes the combined 
efficiencies of the vertex and vertex-plus-mass constrained fits
of $0.952\pm0.006$, which cancels out in our 
subsequent analysis.  
The quantities
$\epsilon_{geom}$\ are the geometrical efficiencies for finding the
daughter mesons in the tracking fiducial volume,
having the decay exceed the minimum \Pt\ requirements 
on the meson and $B$\ systems  given a $J/\psi$\ candidate, and 
having the $B$\ candidate satisfy the constrained fit requirements.
The quantities
$\epsilon_{c\tau}$\ are the efficiencies of the proper decay length
requirement on the $B$\ candidate in the different decay modes.
The quantities
$\epsilon_{K}$,
$\epsilon_{K_s}$,
$\epsilon_{\pi}$\ and
$\epsilon_{\phi}$\  are the efficiencies for reconstructing the 
$K^+$, $\mKshort$, $\pi^-$\ and $\phi$\ mesons using the charged track
information.

In addition to these efficiencies, we correct the 
observed event rates for
the relevant branching fractions into intermediate states,
$\mBR(K^\circ \rightarrow \mKshort \rightarrow\pi^+\pi^-)=0.3430\pm0.0014$\ and 
$\mBR(\phi(1020)\rightarrow K^+K^-)=0.491\pm0.009$, both taken from
Ref.~\cite{ref: PDG}, and
the isospin weighting factors 
$\mBR(K^{\ast}(892)^0\rightarrow K^+\pi^-)=
 \mBR(K^{\ast}(892)^+\rightarrow K^\circ \pi^+)=
{2/3}$.

Since we are only interested in ratios of efficiencies, the common
terms in these efficiencies cancel, reducing the overall uncertainties.
These include the term $\epsilon_{J/\psi}$, and the reconstruction
efficiencies $\epsilon_K$\ or $\epsilon_{K_s}$\ when they appear in both
the numerator and denominator of the ratio.
A number of other
quantities do not necessarily cancel when calculating the ratio of
branching fractions.  In order to evaluate the relative efficiencies, we
employed a $B$\ meson Monte Carlo calculation. 
$B$\ mesons were generated with a 
\Pt\ spectrum predicted by a next-to-leading order QCD 
calculation\cite{ref: NDE spectrum}\ using the MRS~D0 parton
distribution functions \cite{ref: MRS D0}.
The $b$\ quark \Pt\ was required to be $>5$~\GeVc, and the $b$\ quark
fragmentation into a $B$\ meson was modeled using the Peterson parametrization
with the parameter $\epsilon$\ chosen to be 0.006 \cite{ref: Peterson}.
The $B$\ mesons were decayed using the CLEO $B$\ decay 
model\cite{ref: CLEO Monte Carlo}\ and a full simulation was used to model
the response of the CDF detector, including effects due to the underlying
event. 
The resulting Monte Carlo events were then processed with the same algorithms
used to reconstruct the data.
We used the reconstructed Monte Carlo events
 to estimate the geometrical acceptances $\epsilon_{geom}$. 
These efficiencies are listed in Table~\ref{tab: efficiencies}.

The geometrical efficiencies include the
effect of the $B$\ meson vertex and mass constrained fits.  We
determined that the efficiency of the fitting procedure and subsequent
confidence level requirements were independent of $B$\ decay mode
by measuring the relative loss in signal events when different 
event topologies were fit employing both a vertex constraint and a 
vertex-plus-mass constraint.  We assigned a systematic uncertainty
of 1\%\ in the relative acceptances to this effect, which we estimated
by comparing the relative loss of signal events in the different decay
topologies.
We also investigated the uncertainties associated with the model of
the detector used to measure the reconstruction efficiencies.  
We verified that the detector simulations accurately described the interaction
vertex distributions and detector geometry. 
We then compared the efficiencies determined using the complete detector
simulation to those determined using a parametrized model of the detector.
Based on this comparison, we assigned
a 5\%\ systematic uncertainty on the relative geometrical efficiencies
to take into account any remaining uncertainties in the detector model.

In principle, the $B$\ meson $c\tau$\
requirements have different efficiencies for each channel because of
different momentum and vertex resolutions for the final states
and possible differences in the lifetimes of the $B$\ meson states.
We evaluated the efficiencies $\epsilon_{c\tau}$\
using the world average values for the lifetimes of 
the three $B$\ mesons \cite{ref: PDG}, using tracking detector resolutions
observed in the events in the $B$\ sideband regions.
The results
are listed in Table~\ref{tab: efficiencies}.
We repeated the
efficiency calculation varying the lifetimes by one standard deviation.
The resulting variations in the relative efficiencies were assigned as 
systematic uncertainties.

The meson reconstruction efficiencies take into account the charged track
reconstruction efficiencies, the efficiencies of the \Kshort\ lifetime
cut and the additional constrained fits performed when a \Kshort\ candidate
is in the final state.  The track reconstruction efficiencies were
determined using both a full detector simulation and by embedding 
simulated tracks in real interactions containing $J/\psi$\ candidates.  
The systematic uncertainties in the track reconstruction efficiencies
were determined by varying the embedding techniques.
The loss of $K^\pm$\ mesons due to decays-in-flight was estimated using
the full detector simulation.  Between 4\%\ and 6\%\ of $K^\pm$\ mesons
(depending on the $B$~meson decay mode)
decay within the volume of the CTC, of which approximately 40\%\ are
correctly reconstructed.  We assigned a 3\%\ systematic
uncertainty on $\epsilon_K$\ due to the decays-in-flight correction.  

The \Kshort\ lifetime cut efficiency and the efficiencies of the vertex
and mass constrained fits were
determined by measuring the  loss of signal events in both the proper
lifetime distribution and  the $\pi^+\pi^-$\ invariant mass distributions.
The efficiency of the lifetime cut was determined to be
$0.958\pm0.007$.  The uncertainty represents the difference in
efficiencies determined by estimating the loss of real \Kshort\ decays
using the $\pi^+\pi^-$\ invariant mass and the $c\tau$\ distributions for
the candidate samples.
The fractions of \Kshort\ candidates that satisfied the confidence level
requirements on the vertex and vertex-plus-mass constrained fits were
$0.938\pm0.017$\ and $0.983\pm0.006$, respectively.

A number of additional checks were made to verify that
correlations in efficiencies were properly taken into account.
The variation of the dimuon
trigger acceptance for the different $B$\ meson final states was determined 
using a Monte Carlo calculation that simulated both the detector 
response and the effect of the trigger. 
This resulted in a negligible uncertainty in the ratio of acceptances.
Variations in the \Pt\ spectrum of the produced $B$\ mesons 
could also result in a change in the relative acceptance.  This
was measured by varying the
renormalization scale and the $b$~quark mass in the Monte Carlo
calculation of this spectrum.  We assigned a systematic
uncertainty on the relative acceptance due to this effect that varies from 
1 to 5\%, depending on the pair of final states being compared.
The polarization of
the vector mesons in the final state also has an effect on the relative
acceptance.  We varied the longitudinal polarization of the 
$K^\ast(892)$\ meson in the $B$~meson rest frame
by $\pm0.10$\ around a nominal value of 0.75 and the $\phi(1020)$\ meson
longitudinal polarization by 
$\pm0.25$\ around the nominal value of 0.50 \cite{ref: B polarization}. 
We assigned the resulting 2.5\%\ 
change in acceptance as the systematic uncertainty due to this
effect.

We expect \Bu, \Bd\ and \Bs\ mesons to be produced both directly and
through the production of excited $B$\ meson states that
decay to the pseudoscalar mesons we observe.
We investigated the effect such resonant production would have on the
relative ratio of efficiencies of the decay modes studied by
performing a Monte Carlo calculation using the PYTHIA program
\cite{ref: PYTHIA}, which models the production and decay
of higher mass $B$\ meson states.  In this calculation, we 
assumed that the relative production of $B$\ mesons with orbital 
angular momentum $L$\ and spin $S$\ was in the ratio 
$0.30:0.53:0.17$\ for 
$L=1$\ and $S=0$\ or 1: $L=0$\ and $S=1$: $L=0$\ and $S=0$\
\cite{ref: LEP B/B* production}.
The change in the ratio of acceptances,
relative to the case where only pseudoscalar meson production
was assumed, varied
from 1 to 4\%, depending on the decay mode considered.  We 
included this as an additional systematic uncertainty on the 
acceptance.

The systematic
uncertainties assigned to the relative efficiencies are
summarized in Table~\ref{tab: systematic uncertainties}.
These were combined in quadrature to determine the total systematic
uncertainty on the relative acceptance for each pair of decay modes
used in this study. 

\section{Results}
We present our results as a matrix of ratios of acceptance-corrected 
rates of $B$\ meson decays into the five channels.
The observed numbers of signal events, listed in 
Table~\ref{tab: efficiencies},  were  corrected by the
detection efficiency for each decay.
When we form the ten possible ratios of these acceptance-corrected event
rates, the $b$\ quark production cross section and the common efficiencies
cancel. The results are listed in Table~\ref{tab:  BR ratios}.
Three of these ratios have also been determined using a different technique
based on the same data set \cite{ref: J/psi Kshort BR}. The values
determined here are in good agreement with these previous results.
Note that the two measurements of these three ratios are
not statistically independent. 
  
The measured quantities are the ratios of the product of 
$b$~quark fragmentation fractions 
and the $B$~meson branching fractions into the specific final state.
Thus our measurements can be written as
\begin{eqnarray}
R^{K^\circ}_{K^+} = 
{{f_d}\over{f_u}}\
{{
\mBR(B^\circ \rightarrow J/\psi K^\circ)
               }\over{
\mBR(B^+\rightarrow J/\psi K^{+})
						}}
&=&    1.15\pm0.27\pm0.09 , \\
R^{K^{\ast+}}_{K^+} = 
{{
 \mBR(B^+ \rightarrow J/\psi K^{\ast+})
               }\over{
 \mBR(B^+\rightarrow J/\psi K^{+})
						}}
&=&    1.92\pm0.60\pm0.17 , \\
R^{K^{\ast\circ}}_{K^+} = 
{{f_d}\over{f_u}}\
{{
\mBR(B^\circ \rightarrow J/\psi K^{\ast\circ})
               }\over{
\mBR(B^+\rightarrow J/\psi K^{+})
						}}
&=&    1.59\pm0.33\pm0.12, \\
R^{K^\phi}_{K^+} = 
{{f_s}\over{f_u}}\
{{
\mBR(B_s^\circ \rightarrow J/\psi \phi)
               }\over{
\mBR(B^+\rightarrow J/\psi K^{+})
						}}
&=&    0.41\pm0.12\pm0.04, \\
R^{K^{\ast+}}_{K^\circ} = 
{{f_u}\over{f_d}}\
{{
\mBR(B^+ \rightarrow J/\psi K^{\ast+})
               }\over{
\mBR(B^\circ \rightarrow J/\psi K^{\circ})
						}}
&=&    1.68\pm0.58\pm0.11 , \\
R^{K^{\ast\circ}}_{K^\circ} = 
{{
 \mBR(B^\circ \rightarrow J/\psi K^{\ast\circ})
               }\over{
 \mBR(B^\circ\rightarrow J/\psi K^{\circ})
						}}
&=&    1.39\pm0.36\pm0.10, \\
R^{K^{\phi}}_{K^\circ} = 
{{f_s}\over{f_d}}\
{{
\mBR(B_s^\circ \rightarrow J/\psi \phi)
               }\over{
\mBR(B^\circ\rightarrow J/\psi K^{\circ})
						}}
&=&    0.35\pm0.12\pm0.03, \\
R^{K^{\ast\circ}}_{K^{\ast+}} = 
{{f_d}\over{f_u}}\
{{
\mBR(B^\circ \rightarrow J/\psi K^{\ast\circ})
               }\over{
\mBR(B^+\rightarrow J/\psi K^{\ast +})
						}}
&=&    0.83\pm0.27\pm0.07, \\
R^{K^{\phi}}_{K^{\ast+}} = 
{{f_s}\over{f_u}}\
{{
\mBR(B_s^\circ \rightarrow J/\psi \phi)
               }\over{
\mBR(B^+\rightarrow J/\psi K^{\ast +})
						}}
&=&    0.21\pm0.08\pm0.02, \\
R^{K^{\phi}}_{K^{\ast\circ}} = 
{{f_s}\over{f_d}}\
{{
\mBR(B_s^\circ \rightarrow J/\psi \phi)
               }\over{
\mBR(B^\circ\rightarrow J/\psi K^{\ast \circ})
						}}
&=&    0.26\pm0.08\pm0.02,
\end{eqnarray}
where the first and second uncertainties are the statistical and
systematic uncertainties, respectively
(henceforth the first and second uncertainties in measured values will
represent the statistical and systematic uncertainties, respectively). 

We can use these data to constrain both the fragmentation 
fractions and the meson branching fractions.  To extract the branching
fractions, we will have to assume certain ratios of fragmentation fractions.
Correspondingly, we will use phenomenological and theoretical predictions for the
ratios of branching fractions to extract the fragmentation fractions.

\subsection{Branching Fractions}
\subsubsection{The $B_s^\circ$\ Branching Fraction}
The ratios of branching fractions
that involve the \Bs\ meson can be used with the world
average values for the \Bu\ and \Bd\ meson branching fractions into the
four other final states \cite{ref: PDG}\
to estimate the product of the ratio of fragmentation fractions,
$f_s/(f_u,f_d)$, times
the branching fraction
$\mBR(B_s^\circ\rightarrow J/\psi \phi)$.  
The world average \Bu\ and \Bd\ branching fractions have 
been measured assuming
that the fragmentation fractions $f_u$\ and $f_d$\ are equal for
$b$\ quarks produced in $\Upsilon(4S)$\ decays, and so
this assumption is implicit in this calculation.

All ratios that involve the \Bs\ can be rewritten in the form
given by the example
\begin{eqnarray} 
f_s \mBR(B_s^\circ\rightarrow J/\psi \phi) = 
      f_u \mBR(B^+\rightarrow J/\psi K^{+}) R^{\phi}_{K^{+}},
\label{eq: BR determination}
\end{eqnarray}
which gives us four different measures of the ratio of fragmentation
fractions and the \Bs\ branching fraction, using the world average values
for the branching fractions \cite{ref: PDG}\ on the right-hand 
side of Eq.~\ref{eq: BR determination}.
With the assumption of equal \Bu\ and \Bd\ fragmentation fractions, we 
form the weighted average of the four estimates to obtain
\begin{eqnarray}
{{f_s}\over{(f_u,f_d)}} \mBR(B_s^\circ\rightarrow J/\psi \phi) =
\left( 0.37\pm0.11\pm0.04 \right)\times 10^{-3}.
\end{eqnarray}

In order to extract $\mBR(B_s^\circ\rightarrow J/\psi \phi)$, we assume
$f_u=f_d$\ and use the value
$f_s = (0.40\pm0.06)\ f_u$.  
This value of $f_s$\ represents the central value of
the range of reported $f_s$\ 
measurements \cite{ref: K production,ref: Delphi fs}, 
and the uncertainty has
been chosen to cover half of the difference between the minimum and
maximum values.
It is also consistent with
the suppression of strange hadrons observed in the production of
light quark hadrons
\cite{ref: UA1 Compilation}.
With these values for the fragmentation fractions, we determine
\begin{eqnarray}
\mBR(B_s^\circ\rightarrow J/\psi \phi) =
\left( 0.93\pm0.28\pm0.10\pm0.14 \right)\times 10^{-3}.
\end{eqnarray}
The first uncertainty is statistical, the second accounts
for the systematic uncertainties associated with the ratio of 
branching fraction measurements and the third is the uncertainty
associated with the value we have taken for $f_s$.

\subsubsection{The \Bu\ and \Bd\ Branching Fractions}
We can also use these data to estimate the 
branching fractions 
$\mBR(B^+\rightarrow J/\psi K^+)$, 
$\mBR(B^\circ\rightarrow J/\psi K^\circ)$, 
$\mBR(B^+\rightarrow J/\psi K^{\ast+})$\ and
$\mBR(B^\circ\rightarrow J/\psi K^{\ast\circ})$\ 
using the world average values for the 
branching fractions, our ratios of branching fractions and 
the assumption that $f_u = f_d$.
For example, for the decay $B^+\rightarrow J/\psi K^{\ast+}$,
we have three separate estimates 
\begin{eqnarray}
\mBR(B^+\rightarrow J/\psi K^{\ast+})&=& 
\mBR(B^+\rightarrow J/\psi K^+) R_{K^+}^{K^{\ast+}}, \\
\mBR(B^+\rightarrow J/\psi K^{\ast+})&=& 
\mBR(B^\circ\rightarrow J/\psi K^\circ) 
 \left( {{f_d}\over{f_u}} \right) R_{K^\circ}^{K^{\ast+}}, \\
\mBR(B^+\rightarrow J/\psi K^{\ast+})&=& 
\mBR(B^\circ\rightarrow J/\psi K^{\ast\circ}) 
 \left( {{f_d}\over{f_u}} \right) {{1}\over{R_{K^{\ast+}}^{K^{\ast\circ}}}}.
\end{eqnarray}
We use for the first factor on the right-hand side of these estimates the
world average values for the branching fractions \cite{ref: PDG}\
and form the weighted average of these three measurements,
thereby reducing the net statistical uncertainty.
Because we employ in this calculation the world averages that
have been determined assuming that $f_u=f_d$, these results
depend implicitly on this assumption. 
Note that this estimate of $\mBR(B^+\rightarrow J/\psi K^{\ast+})$\ is
statistically independent of the world average value for this branching
fraction.

Using this procedure, 
we obtain the branching fractions
\begin{eqnarray}
\mBR(B^+ \rightarrow J/\psi K^+) &=& 
    \left( 0.82\pm0.18 \pm0.07 \right) \times 10^{-3}, \\
\mBR(B^\circ \rightarrow J/\psi K^\circ) &=& 
    \left( 1.14\pm0.27\pm0.09 \right) \times 10^{-3}, \\
\mBR(B^+ \rightarrow J/\psi K^{\ast+}) &=&
    \left( 1.73\pm0.55\pm0.15 \right) \times 10^{-3}, \\
\mBR(B^\circ \rightarrow J/\psi K^{\ast\circ}) &=& 
    \left( 1.39\pm0.32\pm0.11\right) \times 10^{-3}.
\end{eqnarray}
The statistical and systematic uncertainties have been estimated by
weighting the relative contributions in the world average values and our
data. The uncertainties in the world average branching fractions used in
this calculation are
dominated by the most recent measurements by the CLEO 
collaboration \cite{ref: CLEO results}. 
These uncertainties are limited by the size of the CLEO sample, and are
therefore largely statistical and independent.  We have examined the
stability of these estimates to different assumptions concerning the
independence of the quoted systematic uncertainties.  We find that our
results and their estimated uncertainties are insensitive
to possible correlations in the systematic uncertainties in the CLEO
measurements.

\subsubsection{Comparison with Theory}
We have compared our measured ratios of branching fractions times
fragmentation fractions with a calculation of the two-body nonleptonic
decay rates of
$B$\ mesons, using the factorization hypothesis, chiral and heavy quark
symmetries and data from  semileptonic $D$\ meson 
decays \cite{ref: Deandra et al}. We adjusted the predicted ratios by the
world average
$B$\ meson lifetimes \cite{ref: PDG}\ to correct for
the observed lifetime differences of these three states.
Although several recent theoretical calculations
of these branching fractions exist,
we have selected  Ref.~\cite{ref: Deandra et al}\ for this comparison as
it predicts all the branching fractions for the five decays studied here.
The other model calculations have been made with varying
theoretical assumptions and observational constraints, but they generally
predict ratios of branching fractions that are in reasonable agreement
with each other and our observations.
It should be noted that these calculations generally assume the 
validity of factorization as applied to nonleptonic $B$\ meson decays but
they differ in many details, such as the magnitude and shape of 
the form factors for $B$\ meson decay and the experimental
constraints employed in the calculations.
In Ref.~\cite{ref: Deandra et al},
the form factors are normalized to $D$\ meson semileptonic decay data
and are assumed to be consistent with simple pole dominance.
This assumption has been criticised recently 
\cite{ref: Gourdin et al,ref: Aleksan et al}\
in the light of
data on the observed polarization in the decay 
$B\rightarrow J/\psi K^{\ast}$. 

\begin{figure}
\vspace*{6.0in}
\vbox{
\vskip 1.7in
\hbox{
\hskip -1.0in
\includegraphics{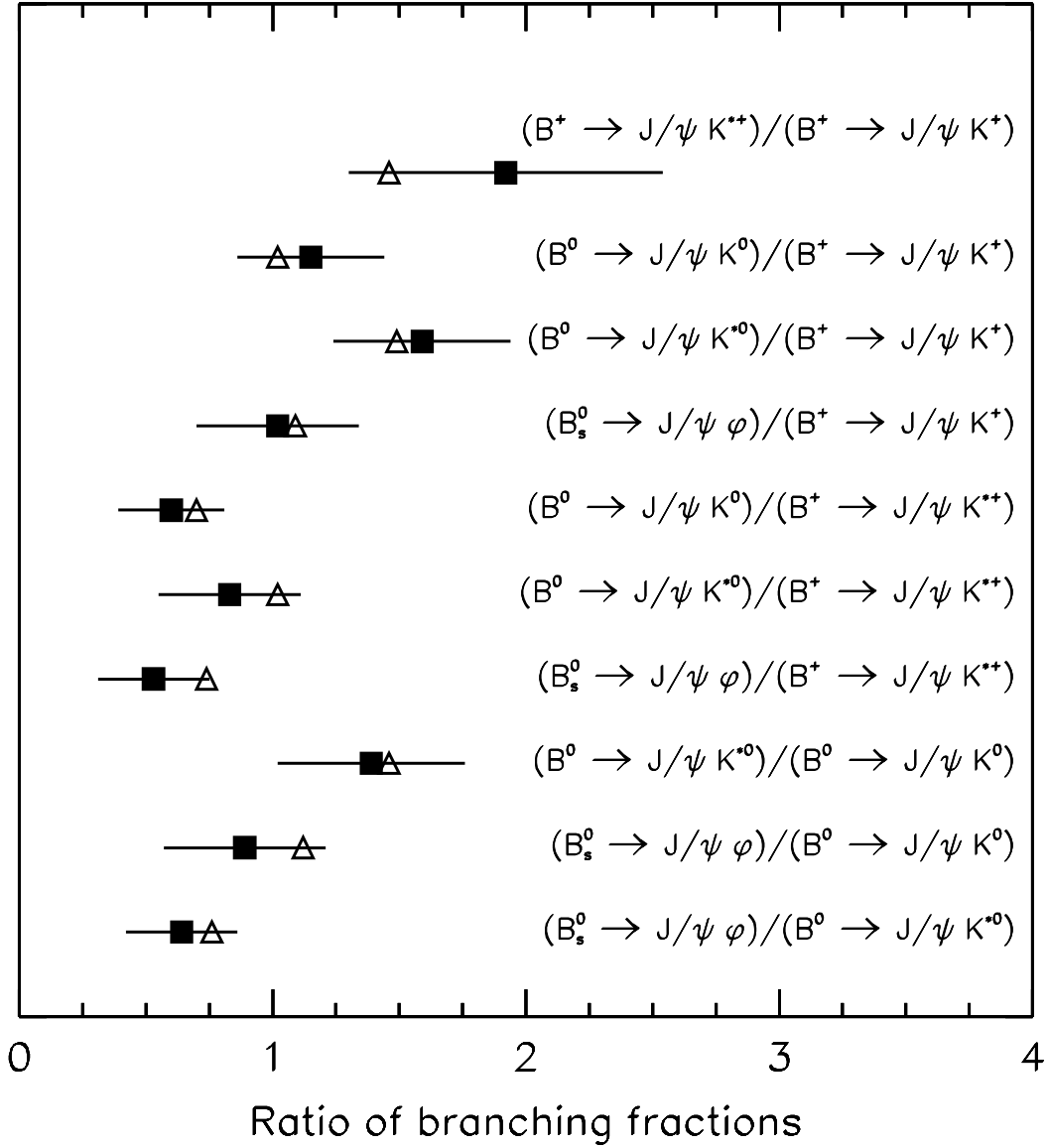}
}
\vskip -2.0in
}
\caption{Comparison of observed ratios of branching fractions (squares) with 
the theoretical prediction (triangles) described in the text.
The error bars reflect the statistical and systematic uncertainties
of the observed ratios added in quadrature.
}
\label{fig: theory cf}
\end{figure}

The results of the comparisons are shown in Fig.~\ref{fig: theory cf}.
In order to compare ratios involving \Bs\ decays, we have assumed 
$f_u=f_d$\ and taken
$f_s = (0.40\pm0.06)\ f_u$.  
The predictions agree well with the observed ratios of
branching fractions for all the decay modes.

\subsection{Ratios of Meson Fragmentation Fractions}
The $b$~quark fragmentation fractions have not been directly measured in
a hadron collider environment.  Our data allow us to constrain the
ratios of these fractions.  However, we note that the measured 
branching fractions of the \Bu\ and \Bd\ mesons have been determined 
assuming equal fragmentation fractions of $b$\ quarks produced
in $\Upsilon(4S)$\ decays.  
We therefore cannot employ the world average values for these 
quantities in estimating $f_u$, $f_d$\ or $f_s$.
Instead, we will make specific assumptions concerning the branching
fractions.

Expressed in terms of fragmentation and branching fractions, 
the two ratios $R_{K^+}^{K^\circ}$\ and $R_{K^{\ast+}}^{K^{\ast\circ}}$\ 
give the relations
\begin{eqnarray}
{{f_d}\over{f_u}} &=& R_{K^+}^{K^\circ} 
           {{\mBR(B^+ \rightarrow J/\psi K^+)}\over{
             \mBR(B^\circ \rightarrow J/\psi K^{\circ})}},
\label{eq: fd over fu 1} \\
{{f_d}\over{f_u}} &=& R_{K^{\ast+}}^{K^{\ast\circ}} 
           {{\mBR(B^+ \rightarrow J/\psi K^{\ast+})}\over{
             \mBR(B^\circ \rightarrow J/\psi K^{\ast\circ})}}.
\label{eq: fd over fu 2}
\end{eqnarray}
Under the assumption that the ratios of branching fractions on the
right-hand side of Eqs.~\ref{eq: fd over fu 1}\ and \ref{eq: fd over fu 2}\
are unity (which is
the result of most quark model predictions), the weighted average
of these two quantities gives
\begin{eqnarray}
{{f_d}\over{f_u}} &=& 0.99 \pm 0.19 \pm 0.08.
\end{eqnarray}
This result is consistent with the hypothesis 
that $b$~quarks hadronize equally often into \Bu\ and \Bd\ mesons.

The strange meson fragmentation fraction $f_s$\ is constrained by the
ratios that involve the \Bs\ final state, yielding the relationships
\begin{eqnarray}
{{f_s}\over{f_u}} &=& R^{\phi}_{K^{\ast+}}
           {{\mBR(B^+       \rightarrow J/\psi K^{\ast+})}\over{
             \mBR(B_s^\circ \rightarrow J/\psi \phi)}}, \\
{{f_s}\over{f_d}} &=& R^{\phi}_{K^{\ast\circ}}
           {{\mBR(B^\circ   \rightarrow J/\psi K^{\ast\circ})}\over{
             \mBR(B_s^\circ \rightarrow J/\psi \phi)}}, \\
{{f_s}\over{f_u}} &=& R^{\phi}_{K^{+}}
           {{\mBR(B^+       \rightarrow J/\psi K^{+})}\over{
             \mBR(B_s^\circ \rightarrow J/\psi \phi)}}, \\
{{f_s}\over{f_d}} &=& R^{\phi}_{K^\circ}
           {{\mBR(B^\circ   \rightarrow J/\psi K^\circ)}\over{
             \mBR(B_s^\circ \rightarrow J/\psi \phi)}}.
\end{eqnarray}
If we assume equal \Bu\ and \Bd\ decay rates to the $J/\psi K^\ast$\ 
final states, equal \Bu\ and \Bd\ decay rates to the 
$J/\psi K$\ final states and $f_u = f_d$, we obtain
the ratios
\begin{eqnarray}
{{f_s}\over{(f_u,f_d)}} &=& \left( 0.24 \pm 0.07 \pm 0.02 \right)
           {{\mBR(B       \rightarrow J/\psi K^{\ast})}\over{
             \mBR(B_s^\circ \rightarrow J/\psi \phi)}}\ {\rm and} 
\label{eq: fs over fud 1}
\\
{{f_s}\over{(f_u,f_d)}} &=& \left( 0.39 \pm 0.11 \pm 0.04 \right)
           {{\mBR(B       \rightarrow J/\psi K)}\over{
             \mBR(B_s^\circ \rightarrow J/\psi \phi)}}.
\label{eq: fs over fud 2}
\end{eqnarray}

The probability of \Bs\ meson production inferred from these
data depends on the ratios of branching fractions 
in Eq.~(\ref{eq: fs over fud 1}) and (\ref{eq: fs over fud 2}).
We take for the ratios of these fractions the values predicted in
\cite{ref: Deandra et al}\ and correct
for lifetime differences as discussed earlier.
We find
that 
\begin{eqnarray}
{{f_s}\over{(f_u,f_d)}} = 0.34\pm0.10\pm0.03.  
\end{eqnarray}

In phenomenological fragmentation models,
the probabilities $f_u$, $f_d$\ and $f_s$\ are related 
to the relative probabilities of producing a $u\bar{u}$, $d\bar{d}$\ and
$s\bar{s}$\ quark pair in the 
quark fragmentation process \cite{ref: fragmentation models}.  
Measurements of the relative probabilities of strange meson to
light meson production
in $e^+e^-$\ and hadron-hadron collisions 
\cite{ref: K production}
and in
deep inelastic scattering have yielded values in the range of 0.3 to 0.4.
A recent compilation of these data has
yielded a
value of $0.29\pm0.015$\ for the relative rate of strange quark production
to up or down quark production \cite{ref: UA1 Compilation}, 
which agrees well with our values
measured in $b$\ quark fragmentation.
Taken together, these measurements indicate that 
the rate of $s\bar{s}$\ suppression in quark fragmentation is 
largely independent of energy and flavor of the quark initiating the
fragmentation process.

\subsection{Fragmentation Fractions of $B$~Hadrons}
The hadronization of $b$\ quarks produces both $B$\ mesons and baryons.
Most $B$\ hadron decay models predict that virtually all $B$\ baryons 
produced during the fragmentation process will subsequently decay via
modes that include a $\Lambda_b^\circ$\ baryon, and so we make that 
assumption here.
A measurement of the rate of $\Lambda_b^\circ$\ production in $b$\ quark
fragmentation gives us
a direct measure of $f_{\Lambda_b}$, the probability that a bottom quark will
hadronize such that a $\Lambda_b^\circ$\ baryon is produced.
This, combined with our measurements of the ratios of fragmentation
fractions, allows us to make a determination of the
values of $f_u$, $f_d$\ and $f_s$.

Studies of $\Lambda_c^+$\ production in semileptonic $b$\ quark 
decays \cite{ref: DELPHI Lambdab,ref: ALEPH Lambdab,ref: OPAL Lambdab}\
have yielded measurements of the product
\begin{eqnarray}
f_{\Lambda_b} \mBR(\Lambda_b^\circ \rightarrow \Lambda_c^+ \ell^- \bar{\nu}_\ell X).
\end{eqnarray}
The na\"\i ve spectator quark model would predict that the 
inclusive $\Lambda_b^\circ$\
semileptonic branching fraction to a final state with a charm
hadron $X_c$\ is in the range of $0.10$.  
Reference~\cite{ref: Mannel and Schuler}\
suggests a possible range of 0.10-0.13.  We have therefore chosen to use the
inclusive branching fraction 
$\mBR(\Lambda_b^\circ \rightarrow X_c \ell^- \bar{\nu}_\ell)
=0.115$, the central value of the theoretical prediction, to estimate 
$f_{\Lambda_b}$.
Assuming this value and the three most recent measurements of $\Lambda_b^\circ$\ 
production in $b$\ quark fragmentation, we find that 
$f_{\Lambda_b} = 0.096\pm0.017$.
It is of interest to note that the observed ratio of meson to baryon
production \cite{ref: UA1 Compilation}\ in minimum 
bias $p\bar{p}$\ collisions at 
$\sqrt{s}=630$~GeV, $6.4\pm1.1$, yields a baryon production fraction
of $0.14\pm0.02$.
This is consistent
with the value determined from $\Lambda_b^\circ$\ semileptonic
decays even though these two fractions are not necessarily expected to
be equal. 

Using the condition that the fragmentation fractions should sum to unity,
assuming that the fraction of charm $B$\ hadrons $f_c\ll 1$\ 
and all $B$\ baryons decay via $\Lambda_b^\circ$\ intermediate states,
then
\begin{eqnarray}
f_u + f_d + f_s + f_{\Lambda_b} = 1.
\end{eqnarray}
This
can be rearranged to determine values for $f_u$, $f_d$\ and $f_s$\
using $f_{\Lambda_b}$\ and our measured ratios of fragmentation fractions:
\begin{eqnarray}
f_u &=& {{1-f_{\Lambda_b}}\over{1 + {{f_d}\over{f_u}} + {{f_s}\over{f_u}} }},\\
f_d &=& {{1-f_{\Lambda_b}}\over{1 + {{f_u}\over{f_d}} + {{f_s}\over{f_d}} }},\\
f_s &=& {{1-f_{\Lambda_b}}\over{1 + {{f_d}\over{f_s}} + {{f_u}\over{f_s}} }}.
\end{eqnarray}
We find that $f_u = 0.39\pm0.04\pm0.04$, 
$f_d=0.38\pm0.04\pm0.04$\ and $f_s=0.13\pm0.03\pm0.01$.
These values are all proportional to the term $(1-f_{\Lambda_b})$\ and
are therefore relatively insensitive to our assumption concerning 
the $\Lambda_b^\circ$\ semileptonic branching fraction.   

\section{Conclusion}
We have measured the ratios of branching fractions times fragmentation fractions
for the five decay modes
$B^+ \rightarrow J/\psi K^+$,
$B^+ \rightarrow J/\psi K^{\ast}(892)^+$,
$B^\circ \rightarrow J/\psi K^\circ$,
$B^\circ \rightarrow J/\psi K^{\ast}(892)^0$\ and
$B_s^\circ \rightarrow J/\psi\phi(1020)$. 

We have used these measurements, with the assumption that 
$f_u = f_d$\ and with $f_s = (0.40\pm0.06) f_u$, to determine the
relative branching fractions of the
$B$\ mesons into the observed final states.  
We have made the first measurement of a \Bs\ branching fraction to 
a final state with a $J/\psi$\ meson, yielding
\begin{eqnarray}
\mBR(B_s^\circ\rightarrow J/\psi \phi) =
\left( 0.93\pm0.28\pm0.10\pm0.14 \right)\times 10^{-3}. \nonumber
\end{eqnarray}
We have also used our data in conjunction with the current world average
branching fractions to find
\begin{eqnarray}
\mBR(B^+ \rightarrow J/\psi K^+) &=& 
    \left( 0.82\pm0.18 \pm0.07 \right) \times 10^{-3}, \nonumber \\
\mBR(B^\circ \rightarrow J/\psi K^\circ) &=& 
    \left( 1.14\pm0.27\pm0.09 \right) \times 10^{-3}, \nonumber \\
\mBR(B^+ \rightarrow J/\psi K^{\ast+}) &=&
    \left( 1.73\pm0.55\pm0.15 \right) \times 10^{-3}, \nonumber \\
\mBR(B^\circ \rightarrow J/\psi K^{\ast\circ}) &=& 
    \left( 1.39\pm0.32\pm0.11\right) \times 10^{-3}. \nonumber
\end{eqnarray}
These data are consistent with a \Bs\ branching fraction approximately
equal to those of the \Bu\ and \Bd\ decays into topologically similar final 
states.
The observed branching fractions are in good agreement with model
calculations employing factorization, chiral symmetry and heavy quark
symmetries.

An analysis of the ratios of branching fractions supports the
widely-held assumption that the probabilities of producing \Bu\ and \Bd\ 
mesons in $b$~quark fragmentation are equal.  We measured the ratio of 
$f_d$\ to $f_u$\ to be $0.99\pm0.19\pm0.08$.
If we assume the theoretically predicted ratios of branching fractions
for the $B_s^0 \rightarrow J/\psi \phi(1020)$\ relative to topologically
similar decay modes for the \Bu\ and \Bd\ mesons and we assume
$f_u=f_d$, we determine that
$f_s/f_u=0.34\pm0.10\pm0.03$.  
Employing an estimate for the fraction of $\Lambda_b^\circ$\ production 
in $b$\ quark fragmentation, $f_{\Lambda_b} = 0.096\pm0.017$, we determine 
$f_u = 0.39\pm0.04\pm0.04$, 
$f_d = 0.38\pm0.04\pm0.04$\ and 
$f_s = 0.13\pm0.03\pm0.01$.
Thus, our results imply a suppression of $s\bar{s}$\ 
production relative to $u\bar{u}$\ and $d\bar{d}$\ production in $b$~quark
fragmentation similar to that measured in $e^+e^-$\ and deep inelastic
scattering experiments.

\section{Acknowledgements}
We thank the Fermilab staff and
the technical staff at the participating institutions for their essential
contributions to this research.
This work is supported by the U.~S.~Department of Energy and the National
Science Foundation; the Natural Sciences and Engineering Research Council
of Canada; the Istituto Nazionale di Fisica Nucleare of Italy;
the Ministry of Education, Science and Culture of Japan;
the National Science Council of the Republic of China;
and the A.~P.~Sloan Foundation.

% now the references. delete or change fake bibitem. delete next three
%   lines and directly read in your .bbl file if you use bibtex.

\newpage

\begin{figure}
\begin{picture}(432,12)
%\put( 50,40){\bf (a)}
%\put(266,40){\bf (b)}
\end{picture}
%\vspace*{1.2in}
\vspace*{1.47in}
\vskip 0.5in
\hbox{
\includegraphics{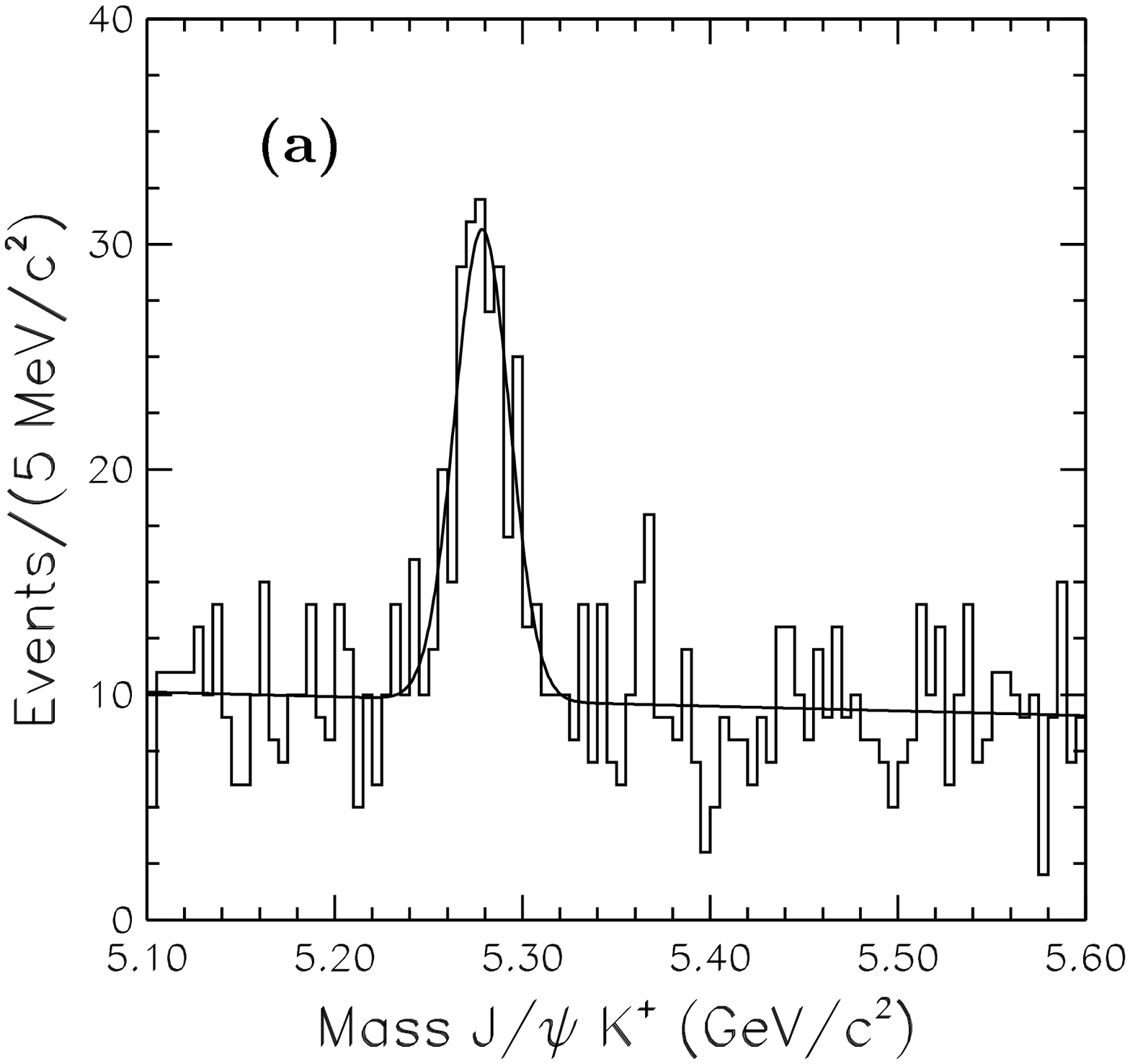}
\hskip 3.0in \includegraphics{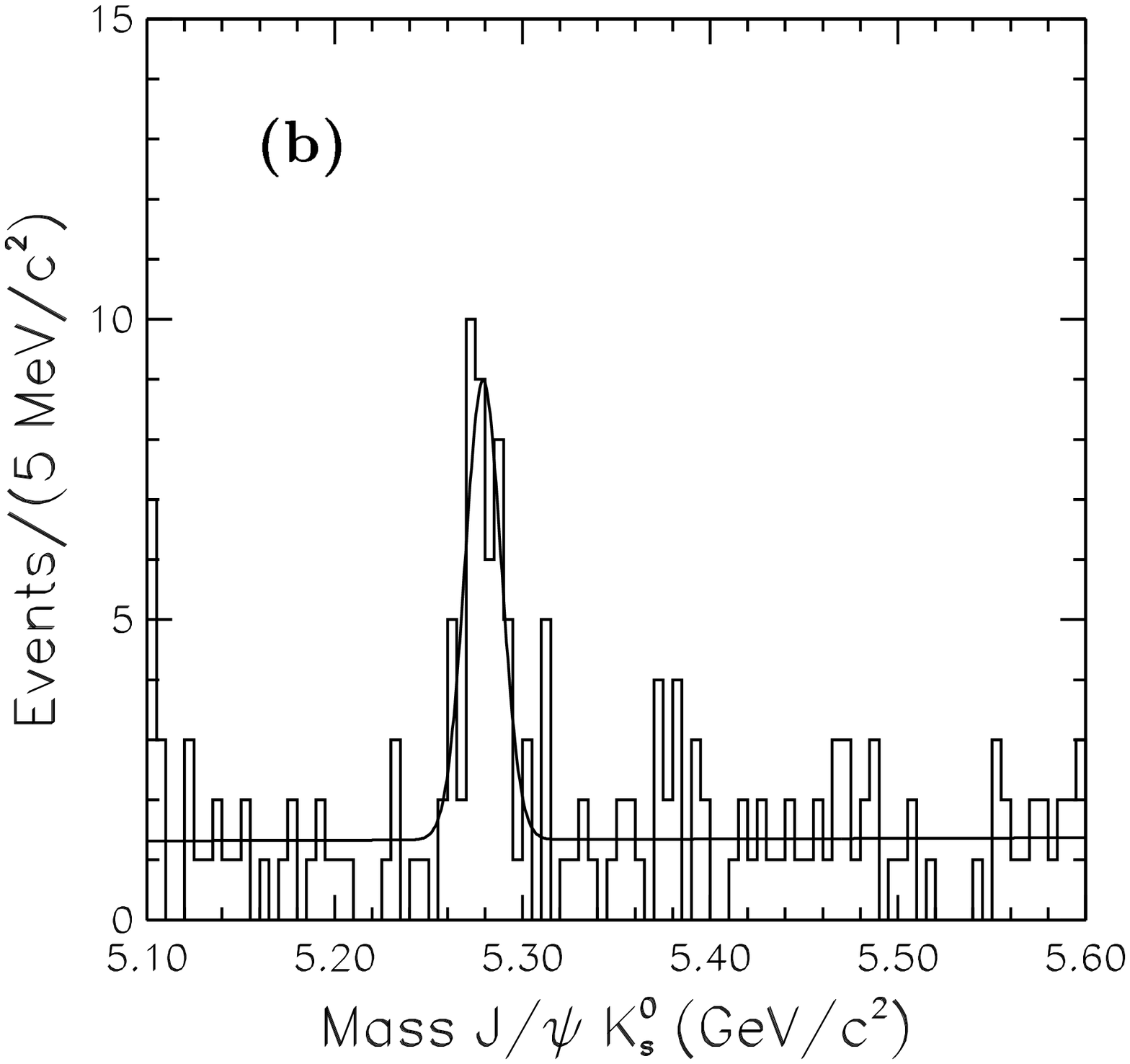}
}
\begin{picture}(432,12)
%\put( 50,10){\bf (c)}
%\put(266,10){\bf (d)}
\end{picture}
\vspace*{2.2in}
\hbox{
\includegraphics{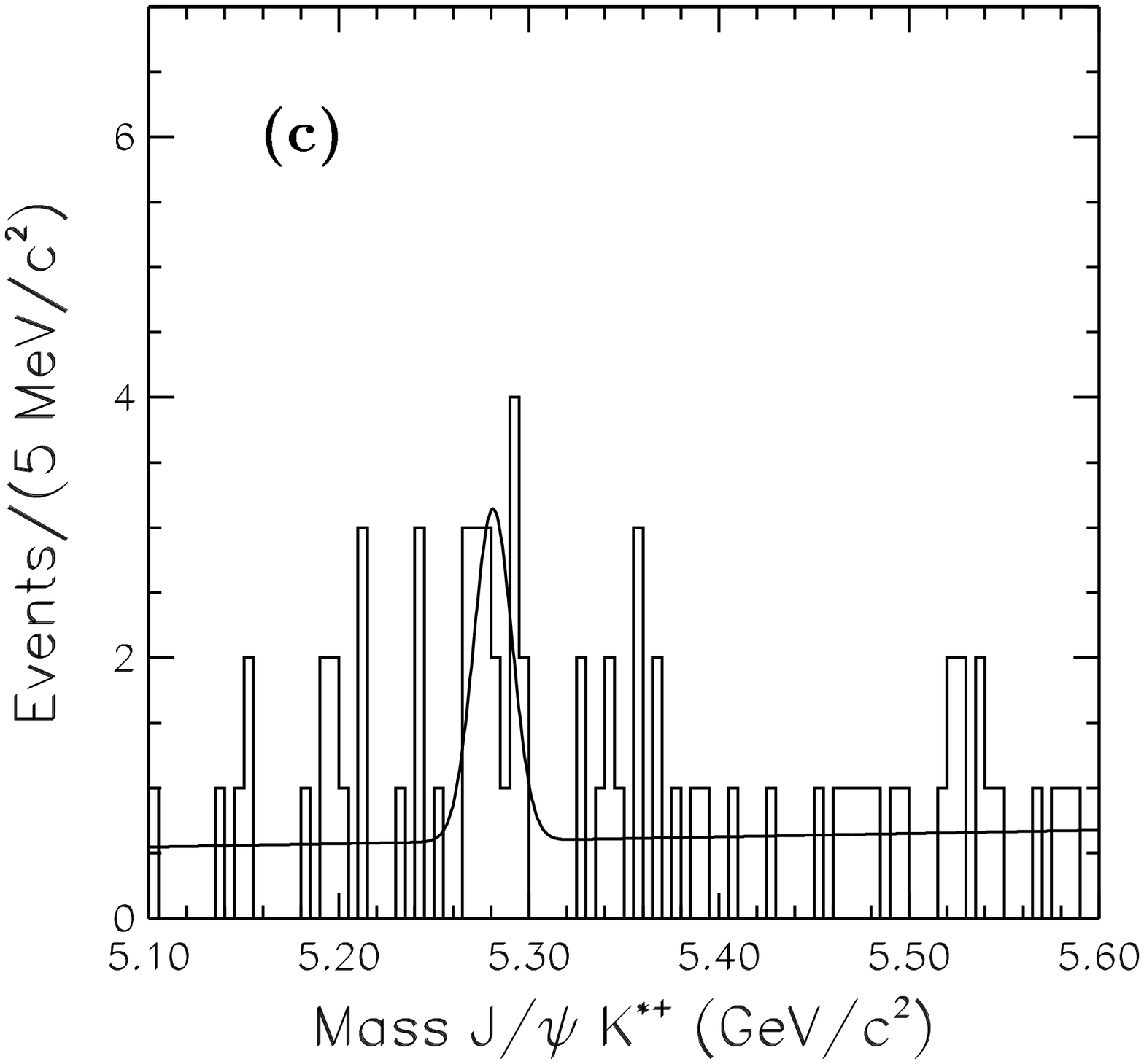}
\hskip 3.0in \includegraphics{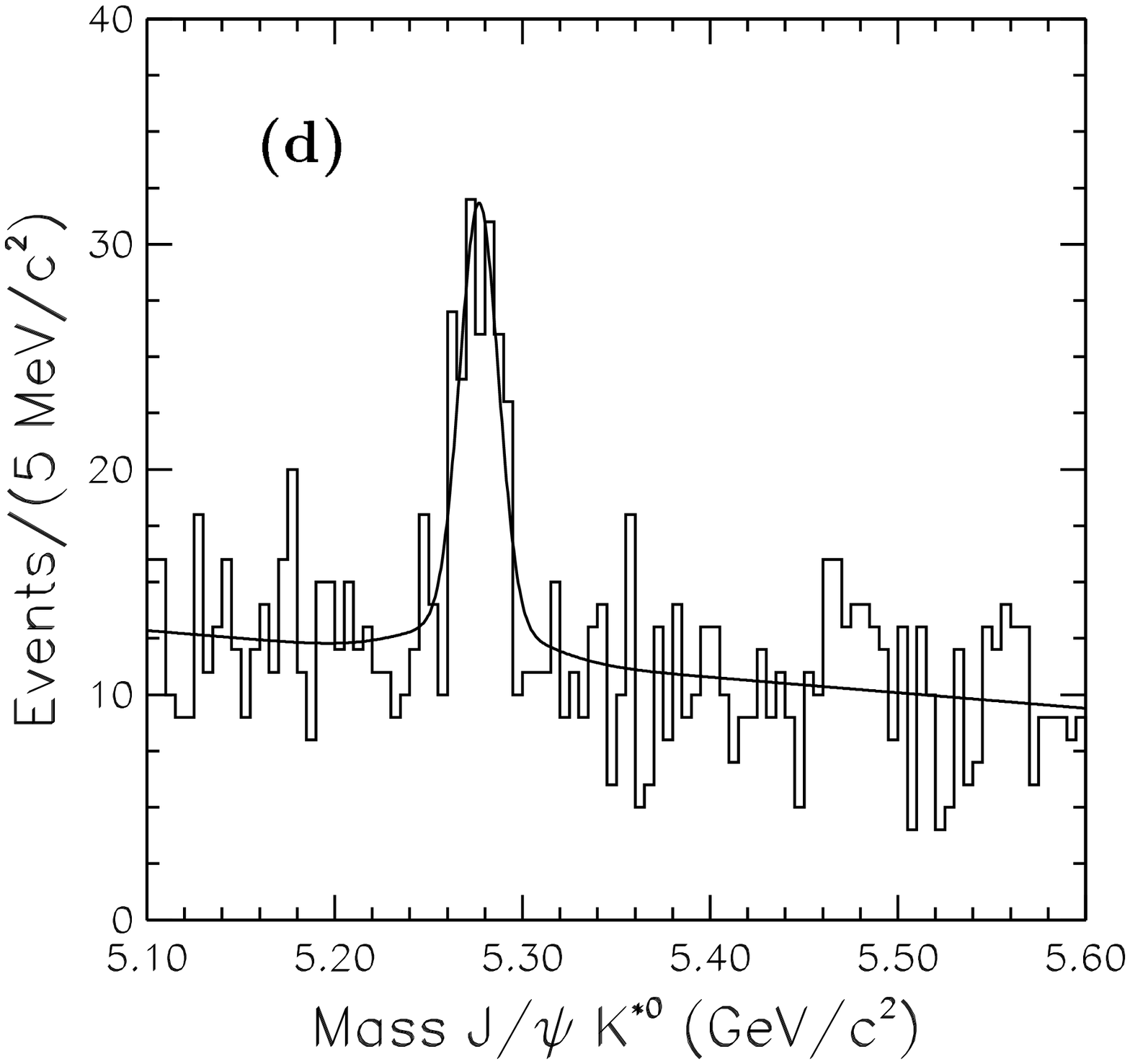}
}
\begin{picture}(432,12)
%\put(170,0){\bf (e)}
\end{picture}
\vspace*{2.2in}
\vskip 0.1in
\hbox{
\hskip 1.5in \includegraphics{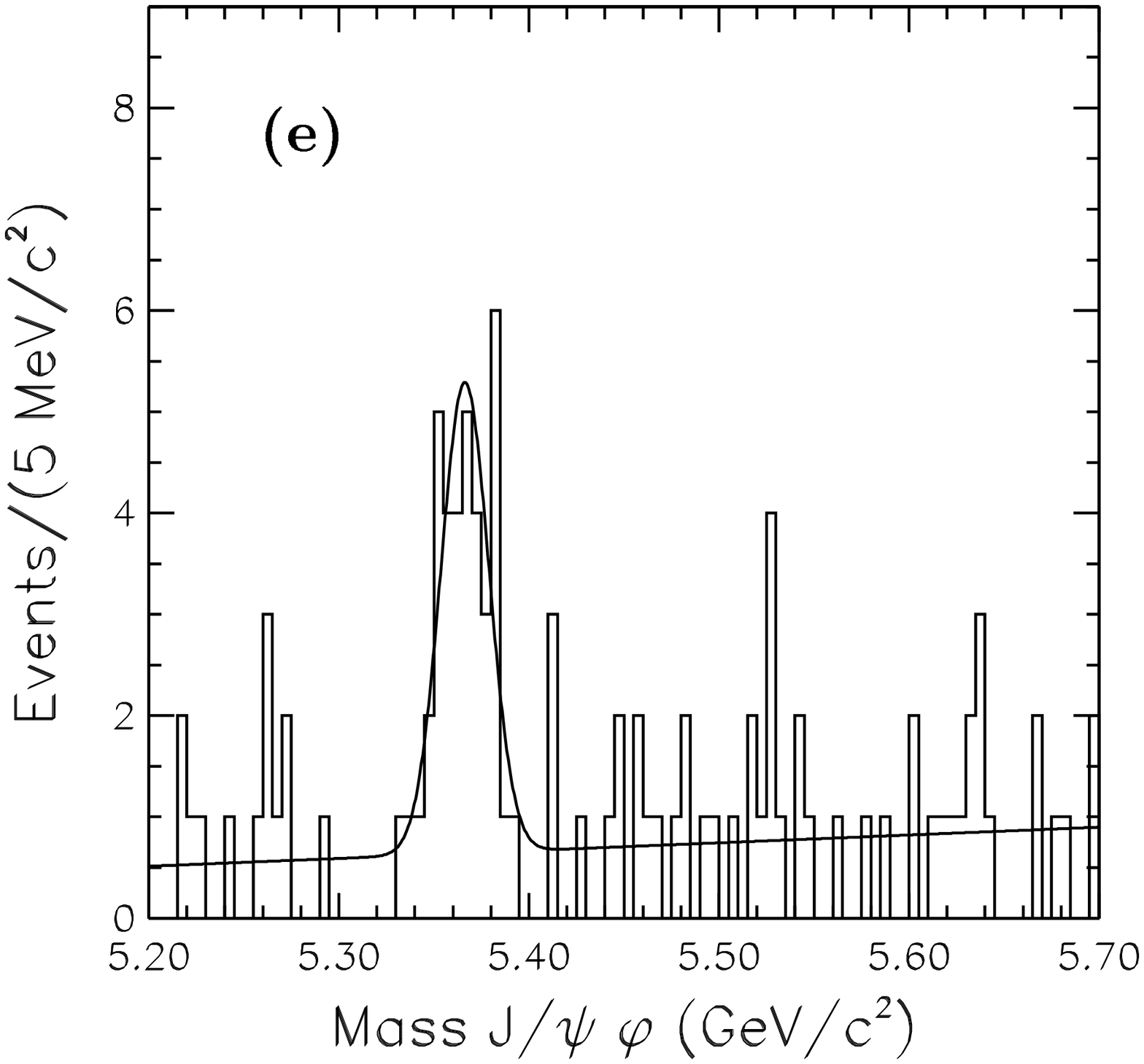}
}
\caption{The $J/\psi K^+$\ invariant mass distribution is shown in (a).  The
invariant mass distributions for the 
$J/\psi K^\circ_s$,  
$J/\psi K^{\ast +}$,
$J/\psi K^{\ast\circ}$\ and
$J/\psi \phi$\ candidate events 
are shown in (b) through (e), respectively.
}
\label{fig:  Jpsi K Mass}
\end{figure}
\newpage

\newpage

\newpage

% \begin{table}
% \caption{}
% \label{}
% \end{tabular}
% \end{table}

\begin{table}
\caption{The definition of the signal regions, sideband regions 
and the number of observed events associated
with the decays $B^+\rightarrow J/\psi K^{\ast +}$,
$B^\circ \rightarrow J/\psi K^{\ast\circ}$\ and
$B_s^\circ \rightarrow J/\psi \phi$.  The number
of observed events is calculated in three ways: fitting the 
resonant structure in the $K\pi$\ and $KK$\ invariant mass distributions
either using the sideband distributions to estimate the background
($N_{sb}$) or using a second order polynomial to estimate the 
background ($N_{fit})$, and using the number of observed $B$\ candidate
events, correcting for the loss of resonant decays due to the two-body
mass requirement ($N_{win}$). The number of events obtained using
the sideband subtracted background, $N_{sb}$, is used to calculate
the branching fraction ratios.
}
\label{tab: sideband regions}
\begin{tabular}{lccc}
		&  $J/\psi \mKshort\pi^+$  &  $J/\psi K^+ \pi^-$  & 
  $J/\psi K^+K^-$ \\ 
\hline
Signal Region (\GeVcc) &
5.235-5.325 &  
5.235-5.325 &  
5.320-5.410 \\
Sideband Regions (\GeVcc) &
5.000-5.220  &  
5.000-5.180  &
5.100-5.305  \\
   &
5.340-5.600  &  
5.380-5.600 &
5.425-5.700 \\
$N_{sb}$\ (events)  &
$21.3\pm6.1$ & $119\pm20$   & $26.7\pm7.3$ \\
$N_{fit}$\ (events)  &
$17.0\pm6.5$ & $108\pm27$   & $27.3\pm7.4$ \\
$N_{win}$\ (events)  &
$16.0\pm5.3$  &  $119\pm18$ & $34.4\pm7.3$ 
\end{tabular}
\end{table}

%\begin{table}
%\caption{The definition of the signal and sideband regions for the
%three-body decays. }
%\label{tab: sideband regions}
%\begin{tabular}{ccc}
%Final State	&	  $B$\ Signal Region (\GeVcc)  &	
%					$B$\ Sideband Region (\GeVcc) \\
%\hline
%$J/\psi \mKshort\pi^+$  &   5.235-5.325	&	5.180-5.220\ {\rm and}\
%5.340-5.600	\\
%$J/\psi K^+ \pi^-$  &	    5.235-5.325	&	5.180-5.220\ {\rm and}\
%5.340-5.600	\\
%$J/\psi K^+K^-$  &	    5.325-5.415	&	5.270-5.310\ {\rm and}\
%5.430-5.470	\\
%\end{tabular}
%\end{table}

%\begin{table}
%\caption{The number of observed $B$\ resonant decays using the
%three techniques described in the text.  The column labelled
%$N_{Sub}$\ gives the number of observed $B$\ decays performing the sideband
%subtraction. The column labelled $N_{fit}$\ gives the number of  observed
%decays fitting the two-body invariant mass distribution to a resonant and
%non-resonant component.
%The column labelled $N_{win}$\ gives the number of observed decays
%assuming no non-resonant contributions in the mass windows used to
%select the $K^\ast(892)$\ and $\phi$\ mesons, and correcting for the
%efficiency of the mass window.
%}
%\label{tab: sideband subtracted rates}
%\begin{tabular}{cccc}
%Channel   &	$N_{Sub}$	&  $N_{fit}$   &  $N_{win}$ \\
%\hline
%$J/\psi \mKshort\pi^+$  & $21.3\pm6.1$ & $17.0\pm6.5$ & $16.0\pm5.3$	\\
%$J/\psi K^+\pi^-$       & $119\pm20$   & $108\pm27$   & $118.8\pm17.8$	\\
%$J/\psi K^+K^-$         & $26.7\pm7.3$ & $27.3\pm7.4$ & $34.4\pm7.3$  \\
%\end{tabular}
%\end{table}

\begin{table}
\caption{The number of observed signal events and the 
reconstruction efficiencies for the five decay modes.
The geometrical efficiency includes the meson \Pt\ requirements and
the acceptance of the tracking fiducial volume. 
It does not include the \Jpsi\ and light quark meson branching fractions.
The efficiencies of the $B$\ proper decay length requirement and the
reconstruction efficiencies of the light quark mesons are also listed.
Some of the systematic uncertainties are correlated as they have common
sources.  These correlations are taken into account when ratios of
the observed decay rates are determined.
}
\label{tab: efficiencies}
\begin{tabular}{ccccc}
Channel	& Events & $\epsilon_{geom}$ &  $\epsilon_{c\tau}$	& $\epsilon_{meson}$ \\
\hline
$B^+\rightarrow J/\psi K^+$ & $154\pm19$ & $(10.1\pm0.8)\times 10^{-2}$	& 
$0.900\pm0.005$	&	 $0.979\pm0.031$ \\
$B^\circ\rightarrow J/\psi K^\circ$ & $36.9\pm7.3$ & $(7.95\pm0.58)\times 10^{-2}$ &
$0.876\pm0.007$ &	 $0.857\pm0.012$ \\
$B^+\rightarrow J/\psi K^{\ast+}$ & $21.3\pm6.1$ & $(4.17\pm0.30)\times 10^{-2}$ &	
$0.880\pm0.006$ &	 $0.839\pm0.015$ \\
$B^\circ\rightarrow J/\psi K^{\ast\circ}$ & $119\pm20$ & $(8.39\pm0.59)\times 10^{-2}$ & 
$0.893\pm0.006$	&        $0.958\pm0.032$ \\
$B_s^\circ\rightarrow J/\psi \phi$ & $26.7\pm7.3$ & $(10.7\pm0.8)\times 10^{-2}$ &	
$0.884\pm0.020$ &        $0.904\pm0.058$ \\
\end{tabular}
\end{table}

\begin{table}
\caption{The systematic uncertainties in the relative efficiencies for
the different channels.
}
\label{tab: systematic uncertainties}
\begin{tabular}{cc}
Effect	&	  Systematic Uncertainty (\%)\\
\hline
$B$\ meson confidence level requirements & 1 \\
Detector simulation 			& 5 \\
$K^+$\ reconstruction efficiency	& 1 \\
Vector meson polarization		& 2.5 \\
\Kshort\ reconstruction			& 1  \\
$B$~\Pt\ spectrum			& 1-5 \\
Effects of excited $B$\ meson production		& 1-4 \\
\end{tabular}
\end{table}

\begin{table}
\caption{The ratios of fragmentation fractions times
branching fractions for the various $B$\
meson final states.  The ratio $R_i^j$\ is located in
the $i$th row and $j$th column.  The uncertainties are statistical and
systematic, respectively.
}
\label{tab:  BR ratios}
\begin{tabular}{ccccc} 
        &       ($J/\psi K^\circ$)   &   
	        ($J/\psi K^{\ast+}$)   &    
	        ($J/\psi K^{\ast\circ}$) &
		($J/\psi \phi$)   \\
\hline
    ($J/\psi K^+$)  
     &  $1.15\pm0.27\pm0.09$ 
     &  $1.92\pm0.60\pm0.17$  
     &  $1.59\pm0.33\pm0.12$
     &	$0.41\pm0.12\pm0.04$ \\  
    ($J/\psi K^\circ$) 
     &  
     &  $1.68\pm0.58\pm0.11$ 
     &	$1.39\pm0.36\pm0.10$
     &	$0.35\pm0.12\pm0.03$ \\
    ($J/\psi K^{\ast+}$) 
     &
     &
     &  $0.83\pm0.27\pm0.07$ 
     &	$0.21\pm0.08\pm0.02$ \\
    ($J/\psi K^{\ast\circ}$) 
     &
     &
     &
     &  $0.26\pm0.08\pm0.02$ \\
\end{tabular}
\end{table}

\end{document}